\documentclass{aa}

\usepackage{natbib}
\bibpunct{(}{)}{;}{a}{}{,} 

\usepackage{graphicx}
\usepackage[varg]{txfonts}
\usepackage{multirow}

\newlength{\digit}
\newcommand{\tc}[1]{\multicolumn{1}{c}{{#1}}}
\newcommand\tstrut{\rule{0pt}{2.5ex}}

\newcommand{\NF}{{$N_\mathrm{frag}\/$ }}

\newcommand{\changed}{}

\begin{document}

   \title{The role of material strength in collisions}

   \subtitle{Comparing solid body and hydrodynamic physics for simulating collisions of planetesimals with icy shells}

   \author{T.~I. Maindl
          \inst{\ref{univie}}
          \and
          R. Dvorak\inst{\ref{univie}}
          \and
          R. Speith\inst{\ref{roland}}
          \and
          C. Sch\"afer\inst{\ref{christoph}}
          }

   \institute{Department of Astrophysics, University of Vienna,
              T\"urkenschanzstra\ss e 17, A-1180 Wien, Austria\\
              \email{thomas.maindl@univie.ac.at, rudolf.dvorak@univie.ac.at}
              \label{univie}
         \and
             Physikalisches Institut, Eberhard Karls Universit\"at T\"ubingen, Auf der Morgenstelle 14, 72076 T\"ubingen, Germany\\
             \email{speith@pit.physik.uni-tuebingen.de}
             \label{roland}
         \and
             Institut f\"ur Astronomie und Astrophysik, Eberhard Karls Universit\"at T\"ubingen, Auf der Morgenstelle 10, 72076 T\"ubingen, Germany\\
             \email{christoph.schaefer@tat.uni-tuebingen.de}
             \label{christoph}
             }

   \date{Received - / Accepted -}

 
  \abstract
   {We investigate the effects of including material strength in multi-material planetesimal collisions.}
   {The differences between strengthless material models and including the full elasto-plastic model for solid bodies with brittle failure and fragmentation when treating collisions of asteroid-sized bodies as they occur frequently in early planetary systems are demonstrated.}
   {\changed{We study impacts of bodies of Ceres-mass with a solid rock impactor and a target with 30\,weight-\% water content as surface ice.} The initial impact velocities and impact parameters are varied between the escape velocity $v_\mathrm{esc}\/$ to about $6\,v_\mathrm{esc}\/$ and from head-on collisions to close fly-bys, respectively. We simulate the collisions using our own SPH code utilizing both strengthless material and the full elasto-plastic material model including brittle failure.}
   {\changed{One of the most prominent differences is the higher degree of fragmentation and shattered debris clouds in the solid model. In most collision scenarios however, the final outcomes are very similar and differ primarily by the about one order of magnitude higher degree of fragmentation in the solid case. Also, the survivors tend to be of less mass in the solid case which also predicts a higher water loss than the strengthless hydro model. This may be an effect of the relatively low-energy impacts that cannot destroy the solid material instantly.  As opposed to giant impacts we also observe an indication that some water ice gets transferred between the bodies.}}
   {}


   \maketitle
%

\section{Introduction}

Existing dynamic studies on the evolution of planetesimals and protoplanets targeting the formation of terrestrial planets assume perfectly inelastic merging \citep[cf.\ ][]{lunobr11} or simplified fragmentation models \citep[e.g.,][]{aleagn98} whenever a collision occurs. By analyzing the bodies' angular momentum \citet{agncan99} show that the assumption of perfectly inelastic merging cannot be sustained as it would lead to rotationally unstable bodies. The true outcome of a collision depends on parameters like the masses involved, collision speed, and the impact angle and can be categorized in one of the four regimes efficient accretion/perfect merging, partial accretion, hit-and-run, and erosion and disruption \citep{asp10}. Accretion efficiency of giant impacts is studied e.g., by \citet{agnasp04} who investigate collision outcomes of two 0.1\,$M_\oplus\/$ bodies with different speeds and impact angles using smoothed particle hydrodynamics (SPH) simulations. \citet{marste09} extend disruption criteria for giant impacts up to a body mass of 10\,$M_\oplus\/$. All in all it is found that 40--50\% of giant collisions are actually not merging events \citep{agnasp04,genkok12,kokgen10}.

\citet{marsas10} show that water contents cannot increase from giant impacts of bodies with masses 0.5\ldots 5\,$M_\oplus\/$. As we are interested in possible water delivery by impacts in early planetary systems we have to look at collision events of smaller bodies, typically involving lower energy and the possibility of water (ice) being transferred between the impactors. Like in the case of planet formation, previous studies simulating the dynamics and collision statistics of asteroid families during and after the Late Heavy Bombardment in the early solar system \citep[e.g.,][]{dvoegg12} assume perfect merging and complete delivery of the asteroids' water content to the impact target. Knowing that this assumption does not hold we need to closely investigate the impact process itself in order to define the conditions under which water is actually transferred rather than being lost during a smaller-scale collision.

For the detailed impact simulations we follow the aforementioned studies on giant collisions and deploy the smooth(ed) particle hydrodynamics (SPH) method, a meshless Lagrangian particle
method developed by \citet{luc77} and \citet{ginmon77} for simulating compressible flows in astrophysical context.
For a detailed description of SPH see, e.g., \citet{mon05,ros09,schspe04}.
Most giant impact studies treat the colliding bodies as strengthless \citep[e.g.,][]{canbar13} by applying the hydro parts of the physical equations defining the material behavior during the impacts. The reasoning is based on the assumption that in giant impacts self-gravity dominates the effects of tensile strength of the material so that hydrodynamics are sufficient to describe the physical effects \citep[e.g., ][]{marste09}. \citet{melrya97} find a size limit of 400\,m in radius for basaltic objects beyond which the energy required to disrupt an asteroid is larger than the energy that is needed for shattering it (i.e., overcoming its tensile strength). Other strengthless models of protoplanet collisions assume a parameter measuring the momentum exchange during the collision which depends on the bodies' material and size \citep[e.g., parameter $\alpha\/$ in ][]{genkok12}. The objects we are dealing with however, have masses of the order of magnitude of $M_\mathrm{Ceres}\/$ and accordingly smaller impact velocities and energies than in the case of giant collisions. Also, we are interested in the details in the collision outcome such as material transfer which is beyond distinguishing between disruption and shattering so that we expect a noticeable contribution of material strength to the simulation outcome.

We use a solid state continuum mechanics model as introduced in SPH by \citet{libpet91}, extended by a model for simulating brittle failure \citep{benasp94,benasp95}, and successfully applied to simulating planetary and asteroid dynamics \citep[cf.\ ][and references therein]{bendur12,micben04,jutasp11}. SPH has also been used successfully for simulating impacts involving agglomerates such as homogeneous protoplanetesimals and comets \citep{gerspe11,jutben08,jutmic09,schspe07}. In this study we focus on identifying the difference the full elasto-plastic model including a damage model for brittle failure and fragmentation makes compared to hydrodynamic treatment of the material when studying collisions of solid, Ceres-sized bodies with water (ice) content.

This paper introduces the two physical models for elasto-plastic continuum mechanics with brittle failure and the hydrodynamic equations in Sect.~\ref{sect:physmod}. In Sect.~\ref{sect:numsim} we briefly discuss our SPH code and describe the simulation scenarios. In  Sect.~\ref{sect:results} we compare the simulation results for the two physical models phenomenologically and conclude in Sect.~\ref{sect:conclusion}, also presenting identified subjects of future more detailed quantitative studies.


\section{The physical model}
\label{sect:physmod}

\subsection{Elasto-plastic materials and brittle failure}
\label{sect:solideq}

For describing the dynamics of solid bodies we use the equations governing the conservation of mass, momentum, and energy formulated according to the theory of continuum mechanics \citep[see for example,][]{schspe07}. For legibility we use the Einstein summation convention and omit particle indices so that the continuity equation in Lagrangian form is
\begin{align}\label{eq:cont}
\frac{\mathrm{d}\rho}{\mathrm{d}t} + \rho \frac{\partial v^\alpha}{\partial x^\alpha} = 0
\end{align}
with the density $\rho\/$ and velocity and positional coordinates $v^\alpha\/$ and $x^\alpha\/$, respectively. In the case of a solid body the stress tensor $\sigma^{\alpha \beta}\/$ takes the role of the pressure $p\/$ and momentum conservation reads
\begin{align}\label{eq:momsolid}
\frac{\mathrm{d}v^\alpha}{\mathrm{d} t} =
  \frac{1}{\rho} \frac{\partial \sigma^{\alpha\beta}}{\partial x^\beta},\quad
  \sigma^{\alpha \beta} = -p\, \delta^{\alpha \beta} + S^{\alpha \beta}
\end{align}
with the deviatoric stress tensor $S^{\alpha \beta}\/$ and the Kronecker delta $\delta^{\alpha \beta}\/$. According to Hooke's law the time evolution of the deviatoric stress tensor is given by the constitutive equations
\begin{align}\label{eq:const}
\frac{\mathrm{d}S^{\alpha \beta}}{\mathrm{d}t} = 2 \mu \left(
\dot{\epsilon}^{\alpha \beta} - \frac{1}{3} \delta^{\alpha \beta}
\dot{\epsilon}^{\gamma\gamma} \right) + S^{\alpha \gamma} R^{\gamma \beta} - R^{\alpha \gamma}S^{\gamma \beta} 
\end{align}
where $\dot{\epsilon}^{\alpha \beta}\/$ is the strain rate tensor
\begin{align}\label{eq:epsdot}
\dot{\epsilon}^{\alpha \beta} = \frac{1}{2} \left( \frac{\partial v^\alpha}{\partial
x^\beta} + \frac{\partial v^\beta}{\partial x^\alpha}\right)
\end{align}
and the rotation rate tensor $R^{\alpha \beta}\/$ is necessary to make the constitutive equations independent from the material frame of reference:
\begin{align}
R^{\alpha \beta} = \frac{1}{2} \left( \frac{\partial v^\alpha}{\partial
x^\beta} - \frac{\partial v^\beta}{\partial x^\alpha} \right).
\end{align}
Conservation of specific inner energy $u\/$ of particles constituting a solid body reads
\begin{align}
\frac{\mathrm{d}u}{\mathrm{d}t} = -\frac{p}{\rho}\frac{\partial v^{\alpha}}{\partial x^\alpha} + \frac{1}{\rho}S^{\alpha \beta}\dot{\epsilon}^{\alpha \beta}.
\end{align}
For closing this set of partial differential equations an equation of state (EOS) connecting pressure, density, and specific energy of the form $p=p\,(\rho,u)\/$ is required. In this work we use the nonlinear Tillotson EOS \citep{til62} as given in \citet{mel89}. It depends on 10 material constants $\rho_0\/$, $A\/$, $B\/$, $a\/$, $b\/$, $\alpha_\mathrm{T}\/$, $\beta_\mathrm{T}\/$, $E_0\/$, $E_\mathrm{iv}\/$, and $E_\mathrm{cv}\/$ and distinguishes three domains: in compressed regions ($\rho>\rho_0\/$) and $u\/$ lower than the incipient vaporization specific energy $E_\mathrm{iv}\/$ the pressure is given by
\begin{align}\label{eq:pl}
p\,(\rho,u) = \left[ a + \frac{b}{1+u/(E_0\,\eta^2)} \right]\rho\,u + A\,\mu_\mathrm{T} + B\,\mu_\mathrm{T}^2
\end{align}
with $\eta = \rho / \rho_0$ and $\mu_\mathrm{T} = \eta-1$. In the expanded state where $u\/$ is greater than the specific energy of complete vaporization $E_\mathrm{cv}\/$ it reads
\begin{align}\label{eq:ph}
p\,(\rho,u) = a\,\rho\,u + & \left[ \frac{b\,\rho\,u}{1+u/(E_0\,\eta^2)}
 + \frac{A\,\mu_\mathrm{T}}{\mathrm{e}^{\,\beta_\mathrm{T}\,(\rho_0/\rho-1)}} \right] \mathrm{e}^{-\alpha_\mathrm{T}\,(\rho_0/\rho-1)^2}.
\end{align}
In the partial vaporization domain i.e., $E_\mathrm{iv}\le u\le E_\mathrm{cv}\/$, the pressure $p\/$ is linearly interpolated between the values obtained via (\ref{eq:pl}) and (\ref{eq:ph}), respectively.

Equations (\ref{eq:cont})--(\ref{eq:ph}) describe the dynamics of a body in the elastic regime. For modeling plastic behavior we apply the von Mises yielding criterion $S^{\alpha \beta}S^{\alpha \beta}>\frac{2}{3}Y^2\/$ \citep[see][]{benasp94} which uses the material yield stress $Y\/$ to limit the deviatoric stress tensor. Hence, we use a transformed deviatoric stress tensor
\begin{align}
S^{\alpha \beta} \rightarrow \frac{2\,Y^2}{3\,S^{\gamma \delta}S^{\gamma \delta}}\, S^{\alpha \beta}
\end{align}
if the von Mises criterion is fulfilled.

We model fracture of the material by adopting the Grady-Kipp fragmentation model \citep{grakip80} following the implementation by \citet{benasp94}. It is based on flaws which are assigned to each SPH particle and which can be activated by a certain strain level $\epsilon\/$. Once active, they develop into cracks and contribute to the damage value $d\in [0,1]\/$ of a SPH particle which is defined as the ratio of cracks to flaws. The deviatoric stress and the tension are reduced proportional to $1-d\/$ and hence vanishes for totally damaged material ($d=1\/$). The initial distribution of the number of flaws $n\/$ with activation threshold $\epsilon\/$ follows a Weibull distribution
\begin{align}
n(\epsilon) = k\,\epsilon^m
\end{align}
with material parameters $k\/$ and $m\/$ \citep{wei39}.

We will use the term \emph{solid model\/} when we refer to the model defined by the equations of this Sect.~\ref{sect:solideq}.

\subsection{The hydro model}
\label{sect:hydroeq}

In order to model strengthless bodies we use the equations of Sect.~\ref{sect:solideq} neglecting deviatoric stress, which eliminates Hooke's law, plastic behavior and the damage model. The equations reduce to
\begin{align}\label{eq:hydro}
\frac{\mathrm{d}\rho}{\mathrm{d}t} + \rho \frac{\partial v^\alpha}{\partial x^\alpha} = 0,\quad
\frac{\mathrm{d}v^\alpha}{\mathrm{d} t} =
  - \frac{1}{\rho} \frac{\partial p}{\partial x^\alpha},\quad
\frac{\mathrm{d}u}{\mathrm{d}t} = -\frac{p}{\rho}\frac{\partial v^{\alpha}}{\partial x^\alpha}
\end{align}
along with the EOS as defined in Eqs.~(\ref{eq:pl}) and~(\ref{eq:ph}). 

We will use the term \emph{hydro model\/} when we refer to the model defined by the equations of this Sect.~\ref{sect:hydroeq}.

Additionally, viscous effects are considered according to the usual SPH artificial viscosity terms in both the solid and hydro models.

\section{Numerical simulations}
\label{sect:numsim}

All simulations are performed with our own 3D parallel SPH code as introduced in \citet{maisch13} with further improvements of performance. The simulations include self-gravity which is needed as we are interested in the global outcome of collisions of bodies of comparable size as opposed to damage done to a target by a high-velocity small body. \changed{This involves adding a term for the gravitational interactions in (\ref{eq:momsolid}) and the corresponding equation in (\ref{eq:hydro}).} We ensure first order consistency and angular momentum conservation by correcting the rotation rate and strain rate tensors as described in \citet{schspe07}.

   \begin{figure}
   \centering
   \includegraphics[width=0.6\hsize]{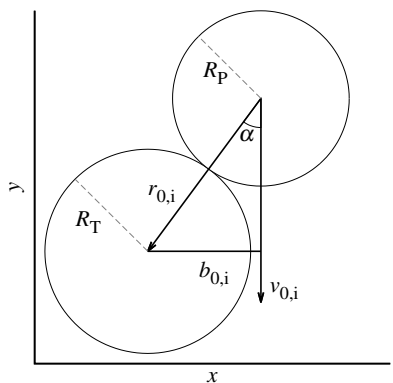}
      \caption{Collision geometry schematic in the target's rest frame. The centers of the two bodies are in the $xy\/$-plane; the impact velocity vector $\vec{v}_{0,\mathrm{i}}\/$ is parallel to the $y\/$-axis and the impact speed is $v_{0,\mathrm{i}}=|\vec{v}_{0,\mathrm{i}}|$. The impact angle $\alpha\/$ is defined as the angle between $\vec{v}_{0,\mathrm{i}}\/$ and the vector $\vec{r}_{0,\mathrm{i}}\/$ connecting the two centers upon impact ($\cos \alpha = \left|\frac{\vec{r}_{0,\mathrm{i}}\cdot \vec{v}_{0,\mathrm{i}}}{r_{0,\mathrm{i}}\, v_{0,\mathrm{i}}}\right|$, $r_{0,\mathrm{i}}=|\vec{r}_{0,\mathrm{i}}|$) and related to the impact parameter upon collision by $b_{0,\mathrm{i}}=r_{0,\mathrm{i}}\,\sin \alpha\/$.}
         \label{fig:configuration}
   \end{figure}
The collisions involve spherical targets and projectiles composed of basalt and water ice. Because of the bodies' spherical symmetry the collisions can be described by the impact angle $\alpha\/$ or equivalently, the impact parameter $b_{0,\mathrm{i}}\/$ and the impact velocity $v_{0,\mathrm{i}}\/$ as illustrated in Fig.~\ref{fig:configuration} along with the respective projectile and target radii $R_\mathrm{P}$ and $R_\mathrm{T}$. Initially neither of the bodies is rotating.

\begin{table*}
\caption{Tillotson EOS parameters, shear modulus $\mu$, and yield stress $Y$ in SI units \citep{benasp99}. Note that $A\/$ and $B\/$ are set equal to the bulk modulus.}
\label{t:EOSpar}
\centering
\begin{tabular}{l*{10}{r}rl}
\hline\hline
\multirow{2}{*}{Material} \tstrut & \tc{$\rho_0$} & \tc{$A$} & \tc{$B$} & \tc{$E_0$} & \tc{$E_\mathrm{iv}$}
         & \tc{$E_\mathrm{cv}$} & \tc{\multirow{2}{*}{$a$}}
                                & \tc{\multirow{2}{*}{$b$}}
                                & \tc{\multirow{2}{*}{$\alpha_\mathrm{T}$}} 
                                & \tc{\multirow{2}{*}{$\beta_\mathrm{T}$}}
                                & \tc{$\mu$} & \tc{$Y$} \\
       & [kg\,m$^{-3}$] & [GPa] & [GPa] & [MJ\,kg$^{-1}$] & [MJ\,kg$^{-1}$] & [MJ\,kg$^{-1}$] & & & & 
       & [GPa]      & [GPa] \\
\hline
Basalt & 2700 & 26.7\hspace{\digit}  & 26.7\hspace{\digit}  & 487 & 4.72\hspace{\digit}  & 18.2\hspace{\digit}  & 0.5 & 1.50 &  5.0 & \tstrut 5.0
& 22.7       & \; 3.5 \\
Ice    & \hspace{\digit}917 & 9.47 & 9.47 & \hspace{\digit}10 & 0.773 & 3.04 & 0.3 & 0.1\hspace{\digit}  & 10.0 & 5.0 & 2.8        & \; 1 \\
\hline
\end{tabular}
\end{table*}
For the Tillotson EOS parameters as well as the shear and bulk moduli we adopt the values given in \citet{benasp99} as summarized in Table~\ref{t:EOSpar}. Following the reasoning in \citet{maisch13} the Weibull distribution parameters of basalt were set to measured values of \citet{nakmic07} and for ice we use those mentioned in \citet{lanahr84}, see Table~\ref{t:weibull}.
\begin{table}
\centering
\caption[]{\label{t:weibull}Weibull distribution parameters.}
\begin{tabular}{llll}
\hline \hline
Material & $m$ & $k\;[\mathrm{m}^{-3}]$ & Reference\tstrut\\
\hline
Basalt   & 16  & $10^{61}$ & (1)\tstrut\\
Ice      & 9.1 & $10^{46}$ & (2)\\
\hline
\end{tabular}
\tablebib{(1)~\citet{nakmic07}; (2) \citet{lanahr84}.}
\end{table}

The masses of the projectile and target $M_\mathrm{P}\/$ and $M_\mathrm{T}\/$, respectively are assumed to be equal and are fixed to Ceres' mass. We use $M_\mathrm{Ceres}=4.74\times 10^{-10}\,M_\odot=9.43\times 10^{20}\,\mathrm{kg}$ which is consistent with \citet{baeche08}. While the \changed{projectile is a solid basalt body the target has a basalt core and a shell of water ice that amounts for $C_\mathrm{T}=30\,\%$ of its mass.} This will allow us to study possible water transfer by the collision. \changed{Accordingly, the target's radius is given by
\begin{align}
\label{eq:RP}
R_\mathrm{T}=\sqrt[3]{\left[C_\mathrm{T}+(1-C_\mathrm{T})\,\frac{\rho_\mathrm{i}}{\rho_\mathrm{b}}\right]\,M_\mathrm{T}\,\frac{3}{4\pi\,}\,\frac{1}{\rho_\mathrm{i}}}
\end{align}
with the respective densities of basalt and ice $\rho_\mathrm{b}\/$ and $\rho_\mathrm{i}\/$. It is somewhat larger than the projectile's radius $R_\mathrm{P}$ (509\,km as opposed to 437\,km).}

In order to cover a large portion of possible collision outcomes for a $M_\mathrm{P}:M_\mathrm{T}=1:1$ mass ratio of projectile and target \citep[see Fig.~10 in][]{leiste12} we arrange our simulations along a grid in initial velocities $v_0\/$ w.~r.~t.\ the target's rest frame and initial impact parameters $b_0\/$. According to typical collision speeds of $M_\mathrm{Ceres}\/$-bodies as determined in \citet{maidvo13} by means of n-body simulations we vary the initial velocities in seven increments covering a range from the target's escape velocity at the initial location of the projectile (231\,m/s) up to 3\,km/s. Six steps were chosen for $b_0\/$ covering central impacts ($b_0=0\/$) up to fly-bys ($b_0=2\,[R_\mathrm{P}+R_\mathrm{T}]=1,892\,\mathrm{km}\/$). Table~\ref{t:initialv} gives the specific initial values that define the parameter grid. The largest $b_0\/$ value allows to study close encounters and to estimate the influence of mutual gravity onto the actual velocities and impact angles.
\begin{table}
\centering
\caption{\label{t:initialv}Initial conditions of the impact simulations. At the beginning the colliding bodies are $r_0=5\,(R_\mathrm{P}+R_\mathrm{T})=4730\,\mathrm{km}$ apart: (a) Initial velocities $v_0$ with normalization by the two-body escape velocity at the initial distance $r_0\/$,
(b) Initial impact parameters $b_0\/$ and hypothetical impact angle $\alpha_0$ (neglecting gravitational interaction, for information only).}
\mbox{
\begin{tabular}{ll}
(a)\\
\hline\hline
\multicolumn{1}{c}{$v_0$} & \multirow{2}{*}{\Large $\frac{v_0}{v_\mathrm{esc}(r_0)}$} \tstrut\\
{}[m\,s$^{-1}$] & \\
\hline
 \hspace{\digit}231 & \hspace{\digit}1\tstrut \\
 \hspace{\digit}516\tablefootmark{a} & \hspace{\digit}2.23 \\
 \hspace{\digit}537\tablefootmark{b} & \hspace{\digit}2.32 \\
1000 & \hspace{\digit}4.33 \\
1500 & \hspace{\digit}6.49 \\
2000 & \hspace{\digit}8.66 \\
3000 & 12.99 \\
\hline
\end{tabular}
}\hfill
\mbox{
\begin{tabular}{rll}
(b)\\
\hline\hline
\multicolumn{1}{c}{$b_0\/$} & Comment\tstrut & \multicolumn{1}{c}{$\alpha_0$}\\
{}[km]  &         & \multicolumn{1}{c}{$[^\circ]$}\\
\hline
     0 & head-on &  \hspace{\digit}0\tstrut\\
   219 & $\frac{1}{2}\,R_\mathrm{T}$ & 13 \\
   437 & $R_\mathrm{T}$ & 28 \\
   819 & - & 60 \\
   946 & $R_\mathrm{P}+R_\mathrm{T}$ & 90 \\
  1892 & $2\,(R_\mathrm{P}+R_\mathrm{T})$ & fly-by \\
\hline
\end{tabular}
}
\tablefoot{
\tablefoottext{a}{Two-body escape velocity upon contact}
\tablefoottext{b}{Target's surface escape velocity}
}
\end{table}
   \begin{figure}
            {\includegraphics[width=\linewidth, clip]{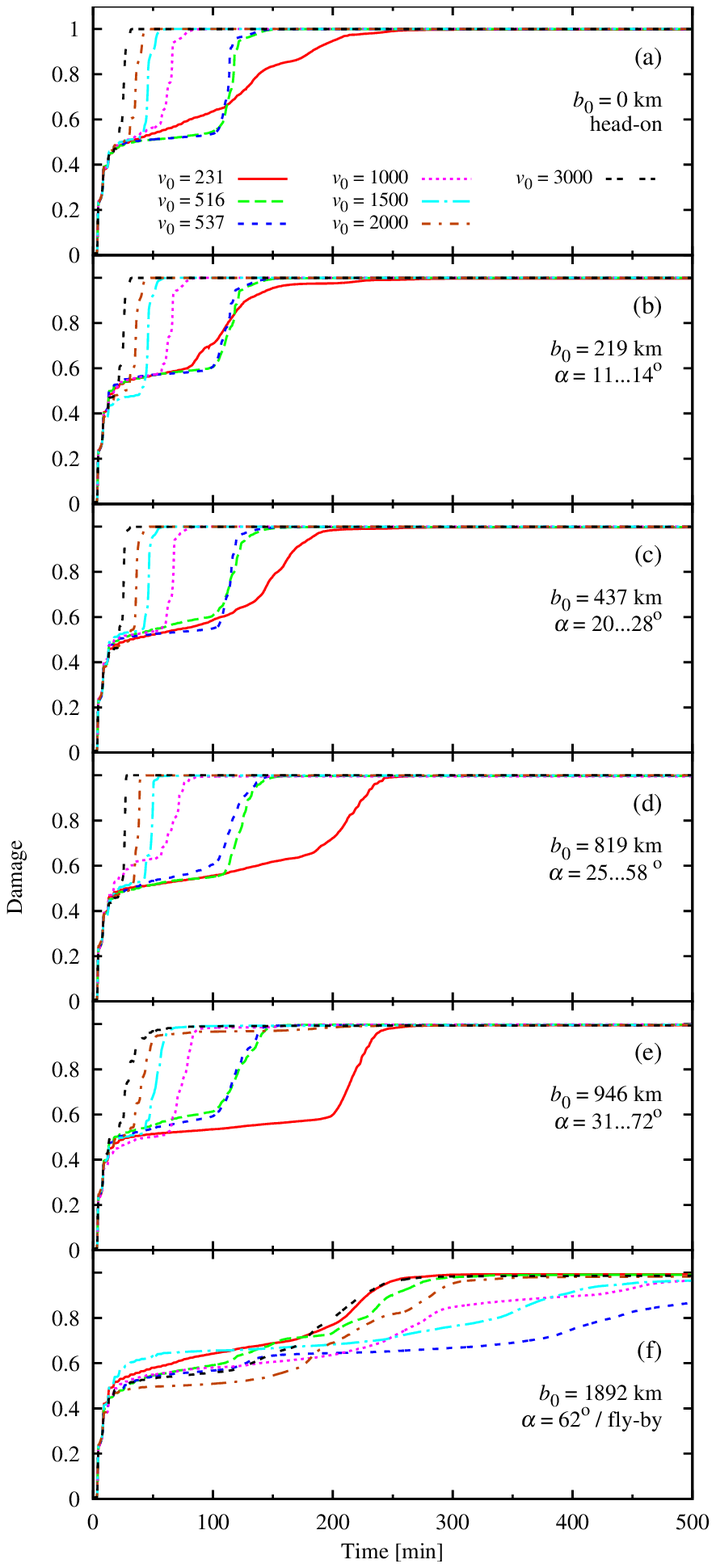}}
      \caption{\changed{Average damage (solid model, cf.~Sect.~\ref{sect:solideq}) per SPH particle until 500\,min into the simulation. Each sub-plot corresponds to one initial impact parameter $b_0\/$ and multiple collision velocities $v_0\/$ as indicated in the top frame. The ranges of the impact angle $\alpha\/$ are taken from Table~\ref{t:actualparameters}. See text for discussion.}
              }
         \label{fig:dam}
   \end{figure}
\changed{The simulations start with the projectile at a central distance of $r_0=5\,(R_\mathrm{P}+R_\mathrm{T})=4730\,\mathrm{km}$ from the target. Hereby we warrant that projectile and target SPH particles do not interact with the respective other body's SPH particles in the beginning. Additionally, the two bodies will approach their respective internal equilibrium before the impact occurs. This can be verified by analyzing how the material damage of the solid model evolves (cf.\ Sect.~\ref{sect:solideq}). Figure~\ref{fig:dam} shows the average damage from simulation start until 500 minutes into the simulation for all scenarios indicated by $b_0\/$ and $v_0\/$. For information we also include the impact angle range for each $b_0\/$ as discussed in Sect.~\ref{sect:actpar} where we also describe fly-by and collision scenarios.
Immediately after starting the simulations an increase in overall damage is noticeable. This is due to internal forces that establish a density gradient primarily driven by self-gravity; in the fly-by scenarios additional tidal forces act upon the bodies during the close encounter contributing to increased damage values. In the colliding scenarios the damage levels off at around 0.5 during the approach which lasts until the collision occurs -- the timespan varies depending on the initial velocity. At the instance of the impact and shortly after, the overall damage quickly rises and gets saturated at a level very close to 1 as illustrated in Fig.~\ref{fig:dam}a--e and by the $v_0=231$\,m/s line in Fig.~\ref{fig:dam}f (colliding scenarios). In case of a near miss ($v_0\ge 516$\,m/s in Fig.~\ref{fig:dam}f) the damage value increases on a longer timescale due to tidal forces between the bodies.
}

\begin{figure*}
   \begin{tabular}{@{}l@{}l@{}}
   \hspace{4em}(a) hydro & \hspace{4em}(b) solid\\[1ex]
   \includegraphics[width=0.5\linewidth, clip]{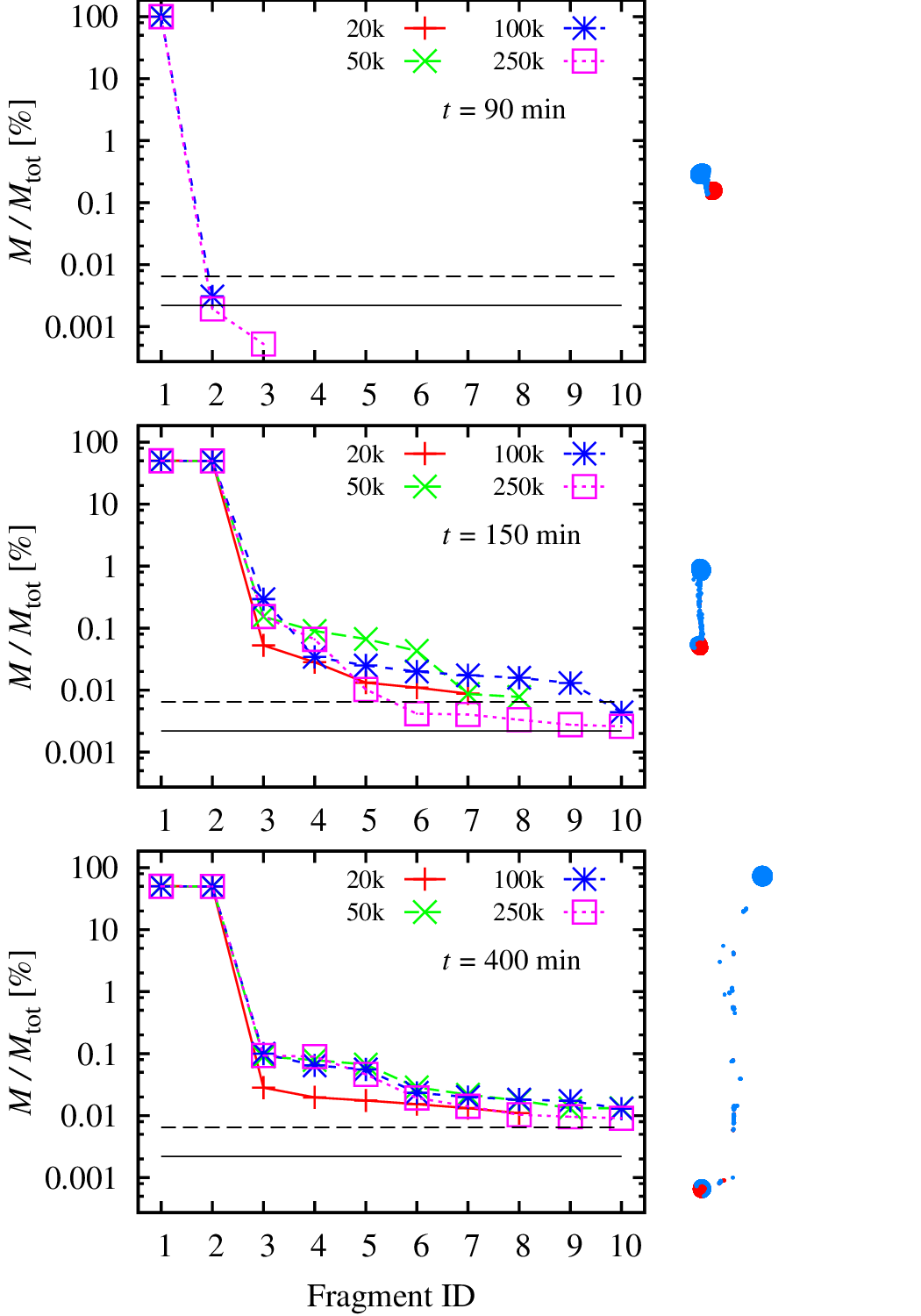} & 
   \includegraphics[width=0.5\linewidth, clip]{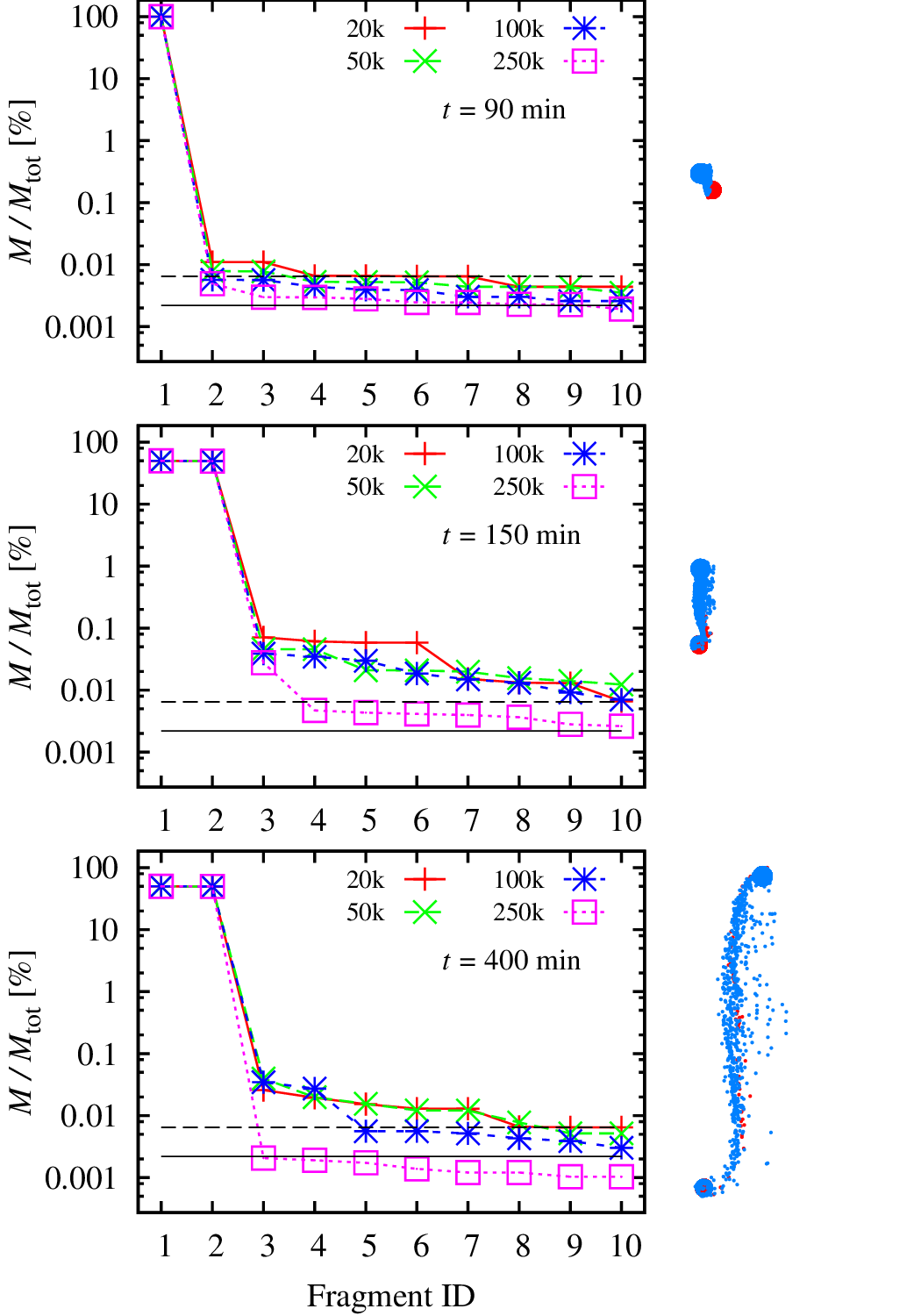}
   \end{tabular}
      \caption{Masses $M\/$ of the 10 largest fragments in the $b_0=946\,\mathrm{km}\/$, $v_0=1\,\mathrm{km\,s^{-1}}\/$ scenario (see underlined values for impact angle and collision velocity in Table~\ref{t:actualparameters}). Masses are expressed as percentages of the total system mass $M_\mathrm{tot}=M_\mathrm{P}+M_\mathrm{T}\/$. We show snapshots at $t=\/$ 90, 150, and 400\,min into the simulation for a total number of SPH particles of 20\,k, 50\,k, 100\,k, and 250\,k, respectively. The underlying physics is (a) the strengthless hydro model and (b) the solid model. To the right of the diagrams we visualize the collision scenario at the corresponding times (resolution 100\,k particles, blue water ice and red basalt particles). Horizontal lines indicate the mass of single basalt (dashed) and ice (solid) particles in the 20\,k-resolution. Note that in the hydro model there are less than 10 fragments in some cases.
              }
         \label{fig:20k-250ki950}
   \end{figure*}

Each of the scenarios defined by $v_0\/$ and $b_0\/$ was simulated over a period of $2,000\,\mathrm{min}\/$ in both the solid and hydro models using an adaptive time step integration scheme in a barycentric frame with output snapshots every 30 seconds resulting in a total of 84 scenarios. For the qualitative investigations we resolve the system in 20,000 SPH particles \changed{(19,986 precisely, due to geometry effects in the initial particle distribution)} following \citet{agnasp04} and the reasoning in \citet{nouemo09}. Also, \citet{genkok12} studied how critical impact velocities depend on the particle numbers and established slight differences for low-resolution scenarios (3,000 particles) and only negligible variations for particle numbers between 20,000 and 100,000.
\changed{To verify our assumption we additionally simulated selected moderate-energy ($v_0=1\,\mathrm{km/s}\/$) impact scenarios with 50\,k, 100\,k, and 250\,k particles, respectively. Via a friends-of-friends algorithm we determine fragments for each snapshot using the smoothing length\footnote{In this study the smoothing length is constant in time and across materials (our choice for the total number of SPH particles results in a separation of approx.\ 35\,km in the initial particle distribution; the smoothing length factor is 2.01).} as the linking length. In order to keep the initial number of interaction partners constant we choose the smoothing length depending on the particle separation in the initial distribution. This will influence the number and size of fragments in different-resolution simulations. Moreover, the fragment mass has a lower limit defined by the mass of a single SPH particle. Hence, in general we expect higher-resolution scenarios to have a smaller lower limit of fragment mass and overall more fragments.
Figure~\ref{fig:20k-250ki950} shows snapshots of the ten biggest fragments' mass-fractions for the four considered SPH resolutions. The scenario has an initial impact parameter $b_0=R_\mathrm{P}+R_\mathrm{T}=946\,\mathrm{km}\/$.
In this case of a (nearly) grazing collision the fragment sizes very rapidly decrease towards the order of one single particle, indicated for the 20\,k particle scenario by the horizontal lines at $M/M_\mathrm{tot}=6.46\times 10^{-5}$ (basalt, dashed) and $M/M_\mathrm{tot}=2.19\times 10^{-5}$ (ice, solid), respectively. Note that even for the 250\,k particle scenario this limit is reached (single-particle mass fractions of $5.17\times 10^{-6}$ and $1.76\times 10^{-6}$, respectively). If we -- for comparisons -- assume a significant fragment to consist of $\gtrsim 20\/$ particles we get a lower mass limit of $\approx 5\times 10^{-4}\,M_\mathrm{tot}\/$.
While, with few exceptions, we observe convergence in the individual
scenarios above this fragment mass limit, the further fragment
distribution differs for different resolutions, in particular for very
small fragments. However, we find the principle differences between
the hydro and solid cases to be independent of the chosen number of
particles. One example is the phenomenon of a significantly larger
degree of fragmentation in the solid scenarios than in the hydro
case. This results in many single-particle fragments which we observe
in all resolutions. As this indicates a part of the object to be
completely destroyed and dispersed, the absolute number of such single
particle fragments of course strongly depends on the resolution. The
higher fragmentation in the solid case will be discussed further in
Sect.~\ref{sect:scenres}. Overall, the results indicate that the chosen resolution
is sufficient to qualitatively study the question at hand,
i.e.\ the fundamental differences between the solid and hydro models
with respect to the collisional outcome.
This will not hold for more detailed collision studies of individual scenarios. In that case several aspects of the methodology need to be tailored to the problem in question: the resolution will have to be higher, the physical model of the bodies (more complex internal structure with multiple material-layers, rotation, etc.), and the code parameters (tree code parameters, time-integration scheme tuning, force calculations, etc.) will have to be adapted. For short-time studies of surface impacts for example, it might prove useful to increase the resolution to 500\,k SPH particles while neglecting gravitation \citep[cf.][]{maidvo14}.

Computation time of our parallel code on contemporary 4 to 8 core CPUs ranges from just under a day to three weeks per scenario depending on the modeled physics, number of interacting SPH particles, relative velocities, etc. Performance obviously also depends strongly on the server's main memory and CPU model. Typically, in our resolution the solid model is computationally more expensive than the hydro model by up to one order of magnitude due to the considerably more complex physics. In case of higher resolutions we expect less influence because the tree code used for self gravity and SPH interaction partner search ($\propto N\,\log N\/$) will dominate the physics ($\propto N\/$).}

\section{Simulation results}
\label{sect:results}

\changed{Throughout the scenarios we observe qualitative differences between the solid and hydro models. Where in the hydro case fragments of the shape of ``bubbles'' are ejected upon impact the solid model resembles the formation of dust-like debris clouds and solid fragments suggesting a significant higher degree of fragmentation. In both cases the fragments themselves seem to accrete debris and grow. The snapshots in Fig.~\ref{fig:20k-250ki950} illustrate this observation for the case of an inclined collision which both bodies survive and subsequently escape.
See Sect.~\ref{sect:scenres} for a systematic description of the scenario results.
}

In analyzing the results the resolution of the method has to be kept in mind. Given the total number of SPH particles resolving the initially homogeneous basalt and water ice along with our choice for the kernel's smoothing length the spatial resolution is about 70\,km.

\subsection{Actual impact parameters and velocities}
\label{sect:actpar}

As the simulation starts with the two bodies set apart by $5\,(R_\mathrm{P}+R_\mathrm{T})=4,730\,\mathrm{km}$ the actual impact angles $\alpha\/$ (or equivalently, impact parameters $b_{0,\mathrm{i}}\/$, cf.\ Fig.~\ref{fig:configuration}) and velocities $v_{0,\mathrm{i}}\/$ will depend on mutual gravitational interaction and tidal forces while the bodies approach each other. \changed{We define the impact to actually happen when the first non-gravitational interaction between at least one projectile and target particle occurs; i.e., when the smallest distance between projectile and target is less than or equal to one smoothing length $h\/$. At this instance we determine the relative position $\vec{r}_{0,\mathrm{i}}\/$ and velocity $\vec{v}_{0,\mathrm{i}}\/$ vectors of the projectile's and target's centers of mass which are effectively mass-weighted averages within the respective target and projectile basalt and ice SPH particles. From $\vec{r}_{0,\mathrm{i}}\/$ and $\vec{v}_{0,\mathrm{i}}\/$ we determine the impact angle $\alpha\/$ (and the impact parameter $b_\mathrm{0,i}\/$).

The time interval $\delta t=30\,\mathrm{s}\/$ between the snapshots used for detecting the collisions and the smoothing length $h\approx 70\,\mathrm{km}\/$ allow us to do an order of magnitude-like estimation of the ``errors'' in $v_\mathrm{0,i}\/$ and $\alpha\/$ introduced by our collision-detection method and the bodies' shapes deviating from spherical symmetry due to tidal forces. Directly from $\delta t\/$ we estimate collision velocity deviations $\delta v \lesssim G\,(M_\mathrm{P}+M_\mathrm{T})/(R_\mathrm{P}+R_\mathrm{T})^2\,\delta t=2.1\,\mathrm{m/s}\/$ ($G\/$ is the gravitational constant). We expect the mean value of the measured mutual barycenter distance to be uncertain by about $\delta r=\pm \max(h/2,v_\mathrm{0,i}\,\delta t/2) \lesssim 45\,\mathrm{km}\/$ with $v_\mathrm{0,i}\approx 3\,\mathrm{km/s}\/$. Additionally, we expect the thus-measured $r_\mathrm{0,i}=|\vec{r}_{0,\mathrm{i}}|\/$ values to be about $h/2\approx 35\,\mathrm{km}\/$ larger than the sum of the bodies' radii.
By using the center of mass distance we assume spherically symmetric bodies at the time of impact and neglect the effects of tidal deformation. To estimate the introduced error we analyzed the measured center of mass distances upon collision $r_\mathrm{0,i,m}\/$ and their deviations from the expected value $R_\mathrm{P}+R_\mathrm{T}+h/2\approx 981\,\mathrm{km}$. The mean and standard deviation are $\overline{r_\mathrm{0,i,m}}=991\,\mathrm{km}\/$ and $\sigma_r=39\,\mathrm{km}\/$, respectively. This is of the order of magnitude of the smoothing length and hence the error induced by assuming spherical symmetry should be consistent with the SPH spatial resolution and the method of determining the first contact of the two bodies. For a rough estimate for the uncertainty of the collision angle $\delta \alpha\/$ when determined via our method we use $r_\mathrm{0,i}\,\sin \alpha=b_\mathrm{0,i}\/$ (cf.\ Fig.~\ref{fig:configuration}). Neglecting a velocity-direction error (which would alter $b_\mathrm{0,i}\/$) we obtain $\delta \alpha = \pm \partial\alpha / \partial r_\mathrm{0,i}\cdot\delta r = \mp \delta r / r_\mathrm{0,i}\,\tan \alpha\/$. Given $\delta r / r_\mathrm{0,i}=\pm 0.05\/$ as determined above, depending on the individual scenario $\delta\alpha\lesssim \pm 9^\circ\/$.}

\begin{table}
\centering
\caption[]{\label{t:actualparsheadon}Actual impact velocities $v_{0,\mathrm{i}}$ for the head-on collision scenarios ($b_0=0$, initial velocity $v_0$), $v_\mathrm{esc}\/$ is the two-body escape velocity upon contact. See text.}
\begin{tabular}{r|*{2}{r}|rr}
\hline \hline
\multicolumn{1}{c}{\tstrut}	& \multicolumn{2}{c}{hydro}		& \multicolumn{2}{c}{solid}		\\
$v_0$	& $v_{0,\mathrm{i}}$	& \multirow{2}{*}{\Large $\frac{v_{0,\mathrm{i}}}{v_\mathrm{esc}}$}	& $v_{0,\mathrm{i}}$	& \multirow{2}{*}{\Large $\frac{v_{0,\mathrm{i}}}{v_\mathrm{esc}}$}	\\
$[\mathrm{m\,s}^{-1}]$	& [m\,s$^{-1}$]	&	& [m\,s$^{-1}$]	&	\\
\hline					
\tstrut 231	 & 492	 & 0.95	 & 496	 & 0.96	\\
					
516	 & 679	 & 1.32	 & 685	 & 1.33	\\
					
537	 & 694	 & 1.34	 & 700	 & 1.36	\\
					
1000	 & 1096	 & 2.12	 & 1096	 & 2.12	\\
					
1500	 & 1566	 & 3.04	 & 1566	 & 3.04	\\
					
2000	 & 2045	 & 3.96	 & 2048	 & 3.97	\\
					
3000	 & 3032	 & 5.88	 & 3032	 & 5.88	\\
\hline
\end{tabular}
\end{table}
\begin{table*}
\centering
\caption[]{\label{t:actualparameters}Actual values for the impact angles $\alpha\pm\delta\alpha\/$ and collision velocities $v_{0,\mathrm{i}}$ for the inclined-collision scenarios defined by initial velocities $v_0$ and initial impact parameters $b_0$. \changed{The collision velocity is also given in terms of the two-body escape velocity upon contact $v_\mathrm{esc}\/$. See text for discussion of the uncertainties $\delta\alpha\/$. The underlined values indicate the scenario shown in Fig.~\ref{fig:20k-250ki950}}}.
{
\begin{tabular}{rl|*{2}{r@{~~~}rr|}*{2}{c@{~~}rr|}c@{~~~}rrr}
\hline \hline
& \multicolumn{1}{l}{\tstrut} & \multicolumn{3}{l}{$b_0$\,[km]}\\
 & & \multicolumn{3}{l|}{219} & \multicolumn{3}{l|}{437} & \multicolumn{3}{l|}{819} & \multicolumn{3}{l|}{946} & \multicolumn{3}{l}{1892}\\
$v_0$ &	\multirow{2}{*}{Model} & \multicolumn{1}{c}{$\alpha$} & $v_{0,\mathrm{i}}$ & \multirow{2}{*}{\Large $\frac{v_{0,\mathrm{i}}}{v_\mathrm{esc}}$}	& \multicolumn{1}{c}{$\alpha$} & $v_{0,\mathrm{i}}$ & \multirow{2}{*}{\Large $\frac{v_{0,\mathrm{i}}}{v_\mathrm{esc}}$}	& $\alpha$ & $v_{0,\mathrm{i}}$ & \multirow{2}{*}{\Large $\frac{v_{0,\mathrm{i}}}{v_\mathrm{esc}}$}	& $\alpha$ & $v_{0,\mathrm{i}}$ & \multirow{2}{*}{\Large $\frac{v_{0,\mathrm{i}}}{v_\mathrm{esc}}$}	& $\alpha$ & $v_{0,\mathrm{i}}$ & \multirow{2}{*}{\Large $\frac{v_{0,\mathrm{i}}}{v_\mathrm{esc}}$}	\\
\small [m\,s$^{-1}$] &	& \multicolumn{1}{c}{\small [$^\circ$]} & \small [m\,s$^{-1}$] &	& \multicolumn{1}{c}{\small [$^\circ$]} & \small [m\,s$^{-1}$] &	& \small [$^\circ$] & \small [m\,s$^{-1}$] &	& \small [$^\circ$] & \small [m\,s$^{-1}$] &	& \small [$^\circ$] & \small [m\,s$^{-1}$] &	\\
\hline						
231 & hydro	 & 6 & 490 & 0.95	 & 11 & 486 & 0.94	 & $21\pm1$ & 486 & 0.94	 & $25\pm1$ & 486 & 0.94	 & 58 & \tstrut488 & 0.95	\\
						
 & solid	 & 14 & 435 & 0.84	 & 23 & 472 & 0.91	 & $25\pm1$ & 513 & 0.99	 & $31\pm2$ & 505 & 0.98	 & 62 & 509 & 0.99	\\[1ex]
						
516 & hydro	 & 9 & 677 & 1.31	 & 19 & 676 & 1.31	 & $37\pm2$ & 675 & 1.31	 & $44\pm3$ & 675 & 1.31	 & - & - & -	\\
						
 & solid	 & 11 & 686 & 1.33	 & 21 & 680 & 1.32	 & $40\pm2$ & 682 & 1.32	 & $48\pm3$ & 681 & 1.32	 & - & - & -	\\[1ex]
						
537 & hydro	 & 9 & 691 & 1.34	 & 19 & 690 & 1.34	 & $38\pm2$ & 690 & 1.34	 & $45\pm3$ & 690 & 1.34	 & - & - & -	\\
						
 & solid	 & 11 & 704 & 1.37	 & 20 & 699 & 1.36	 & $40\pm2$ & 697 & 1.35	 & $48\pm3$ & 694 & 1.34	 & - & - & -	\\[1ex]
						
1000 & hydro	 & 12 & 1094 & 2.12	 & 24 & 1095 & 2.12	 & $48\pm3$ & 1094 & 2.12	 & \underline{$59\pm5$} & \underline{1092} & \underline{2.12}	 & - & - & -	\\
						
 & solid	 & 12 & 1095 & 2.12	 & 25 & 1096 & 2.12	 & $50\pm3$ & 1098 & 2.13	 & \underline{$62\pm5$} & \underline{1095} & \underline{2.12}	 & - & - & -	\\[1ex]
						
1500 & hydro	 & 12 & 1566 & 3.04	 & 25 & 1565 & 3.03	 & $51\pm4$ & 1564 & 3.03	 & $65\pm6$ & 1564 & 3.03	 & - & - & -	\\
						
 & solid	 & 12 & 1566 & 3.03	 & 25 & 1566 & 3.03	 & $53\pm4$ & 1567 & 3.04	 & $67\pm7$ & 1567 & 3.04	 & - & - & -	\\[1ex]
						
2000 & hydro	 & 13 & 2049 & 3.97	 & 27 & 2050 & 3.97	 & $55\pm4$ & 2050 & 3.97	 & $69\pm7$ & 2049 & 3.97	 & - & - & -	\\
						
 & solid	 & 13 & 2050 & 3.97	 & 27 & 2050 & 3.97	 & $55\pm4$ & 2049 & 3.97	 & $72\pm9$ & 2051 & 3.98	 & - & - & -	\\[1ex]
						
3000 & hydro	 & 13 & 3034 & 5.88	 & 28 & 3031 & 5.87	 & $58\pm5$ & 3034 & 5.88	 & $72\pm9$ & 3033 & 5.88	 & - & - & -	\\
						
 & solid	 & 13 & 3033 & 5.88	 & 28 & 3031 & 5.87	 & $58\pm5$ & 3033 & 5.88	 & $72\pm9$ & 3033 & 5.88	 & - & - & -	\\
 
\hline
\multicolumn{2}{c}{\tstrut} &
\multicolumn{3}{c}{$\delta\alpha\lesssim\pm1^\circ$} &
\multicolumn{3}{c}{$\delta\alpha\lesssim\pm1^\circ$} &
\multicolumn{3}{c}{} &
\multicolumn{3}{c}{} &
\multicolumn{3}{c}{$\delta\alpha\lesssim\pm5^\circ$}\\
\hline

\end{tabular}
}
\end{table*}
\changed{Tables~\ref{t:actualparsheadon} and~\ref{t:actualparameters} list the parameters describing the collisions for head-on and inclined impacts, respectively. The actual impact velocites $v_{0,\mathrm{i}}\/$ are also given in units of the two-body escape velocity $v_\mathrm{esc}^2=2\,G\,(M_\mathrm{P}+M_\mathrm{T})/(R_\mathrm{P}+R_\mathrm{T})\/$, $v_\mathrm{esc}=516\,\mathrm{m/s}$. Due to energy dissipation into internal (tidal deformation) energy the collision speed in the $v_0=231\,\mathrm{m/s}\/$-scenario is slightly lower than $v_\mathrm{esc}\/$. In addition to the velocities Table~\ref{t:actualparameters} also states the determined collision angles $\alpha\/$ and their uncertainties $\delta\alpha\/$ as discussed above. As expected for the largest initial offset $b_0=2\,(R_\mathrm{P}+R_\mathrm{T})\/$, mutual gravitational attraction leads to a collision only for small initial velocities $v_0\/$; for $v_0>v_\mathrm{esc}\/$ we observe a fly-by.}

\subsection{Scenario results}
\label{sect:scenres}

    \begin{figure}
    \includegraphics[width=\linewidth, clip]{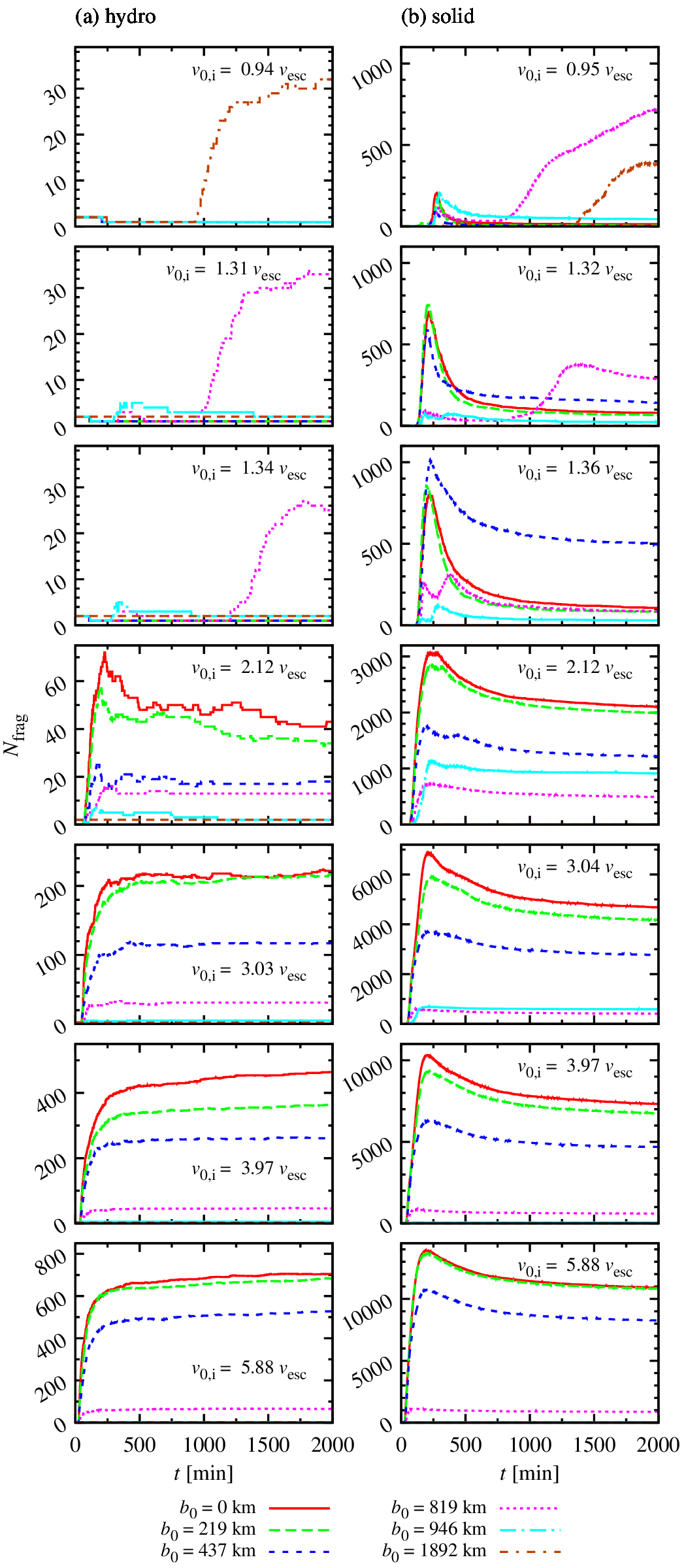}
      \caption{Number of fragments \NF versus time $t\/$ in all the simulated (a) hydro and (b) solid scenarios. The data is organized by initial impact parameter $b_0\/$ and actual collision velocity $v_\mathrm{0,i}\/$. See text for discussion.
              }
         \label{fig:fragcount}
   \end{figure}
\changed{
Figure~\ref{fig:fragcount} shows how the total number of fragments \NF evolves over the course of the simulation (2,000\,min) for (a) the hydro and (b) the solid models. Each frame corresponds to one initial velocity $v_0\/$, the plots state the actual impact velocity $v_\mathrm{0,i}\/$ as averaged from Tables~\ref{t:actualparsheadon} and~\ref{t:actualparameters}. Note that for a given $v_0\/$ the $v_\mathrm{0,i}\/$ variations are $\lesssim 2\,\%\/$. Each individual curve shows the data for one initial impact parameter $b_0\/$ from 0 (head-on) to 1,892\,km (fly-by in most scenarios); see Table~\ref{t:actualparameters} for the individual impact angles $\alpha\/$. The sharp increase of the fragment count -- easier seen in the solid model graphs for low encounter velocities -- indicates the time of collision.

After the collision a large number of fragments is produced and subsequently some of the debris is re-accreted by the surviving bodies as seen by the decline of \NF after the initial peak. This behavior is more prominent in the solid model, especially at higher collision speeds.

The most notable difference between the two models is that in the solid model \NF is more than one order of magnitude greater than in the hydro case. This is caused by an interplay of plasticity and damage model present in the solid but not in the fluid. The material strength makes the bodies more rigid and kinetic energy is partly stored in elastic loading while in the hydro model more kinetic energy results into irreversible deformation. Once the damage threshold is reached, sudden material failure occurs which causes an increase of available energy that cannot contribute to further deformation or enhanced material damage and hence accelerates dust fragments which are not bound by material strength any more. Thus, it is the sudden transition from undamaged to damaged material which releases excess energy leading to higher fragmentation.
Furthermore, during the approach of projectile and target, the fluid experiences larger deformation and therefore higher energy loss due to tidal interactions, eventually leading to a slightly lower actual collision velocity.}

   \begin{figure*}
            \includegraphics[width=\linewidth, clip]{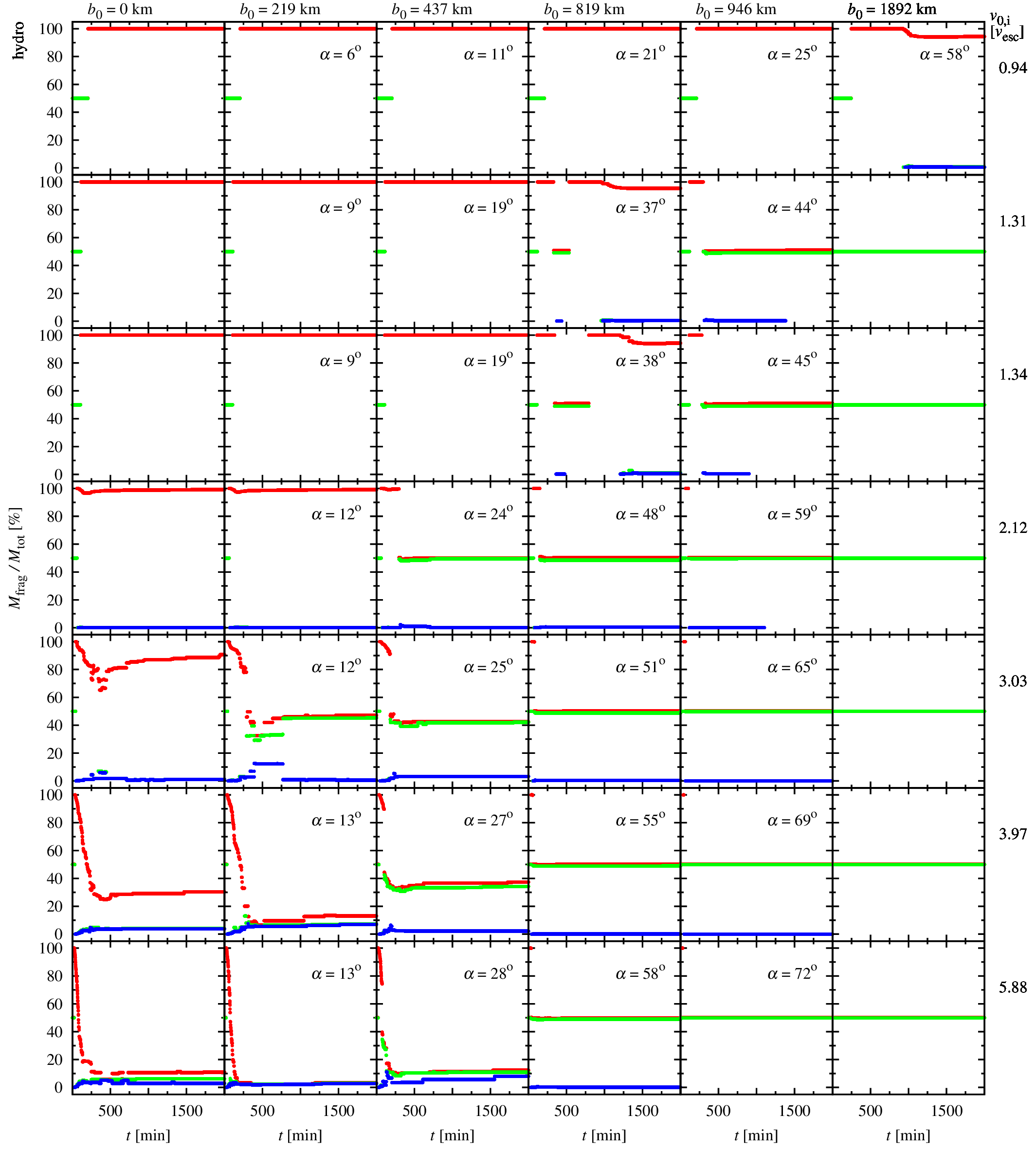}
      \caption{Masses of the three biggest fragments $M_\mathrm{frag}\/$ relative to the total system mass $M_\mathrm{tot}\/$ versus time $t\/$ in all hydro scenarios; $\alpha\/$ is the collision angle for inclined impacts. Initial velocities $v_0\/$ increase row by row, initial impact parameters $b_0\/$ increase from left to right; see text for discussion.
              }
         \label{fig:3biggesth}
   \end{figure*}
   \begin{figure*}
            \includegraphics[width=\linewidth, clip]{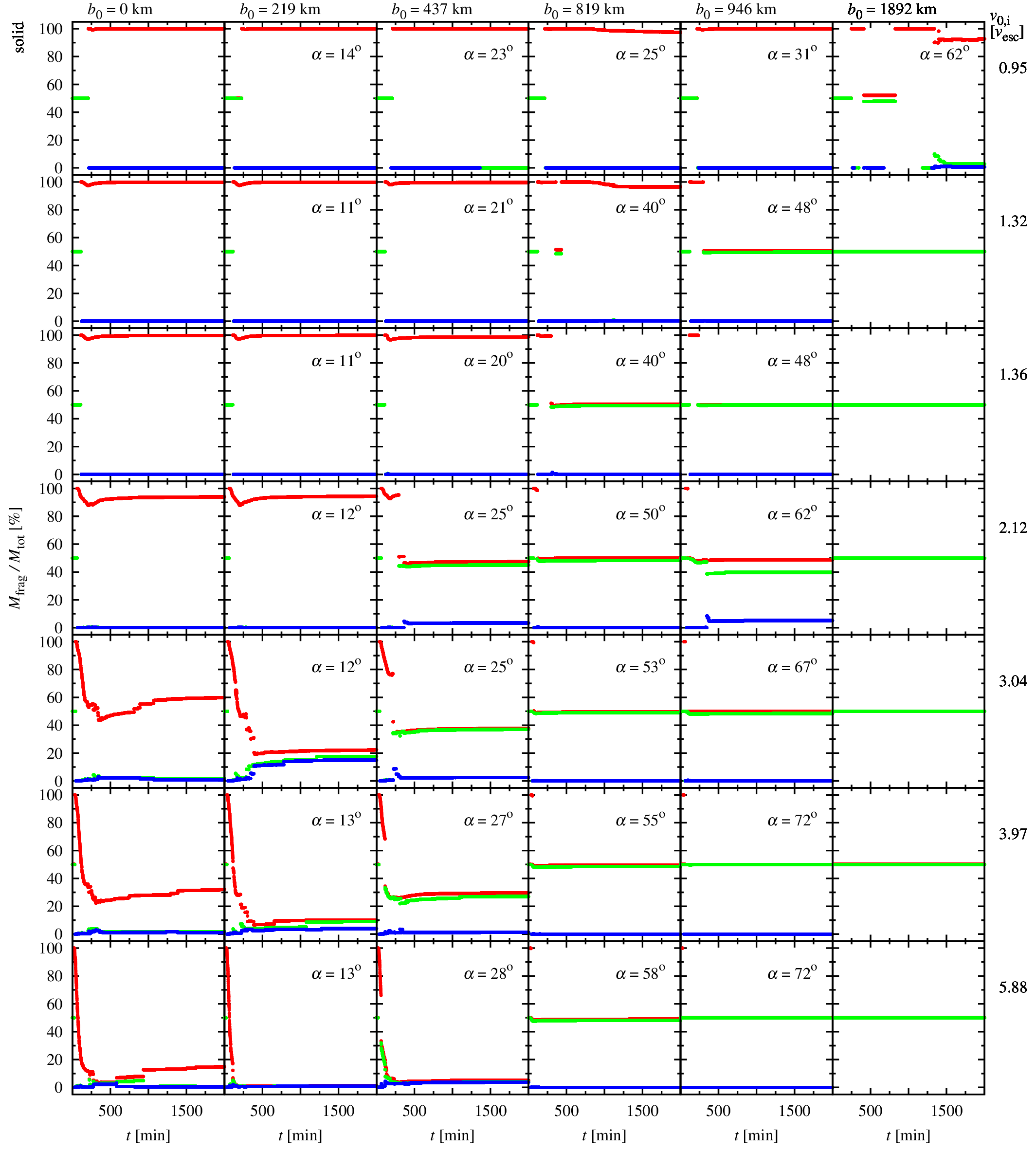}
      \caption{Masses of the three biggest fragments $M_\mathrm{frag}\/$ relative to the total system mass $M_\mathrm{tot}\/$ versus time $t\/$ in all solid scenarios; $\alpha\/$ is the collision angle for inclined impacts. Initial velocities $v_0\/$ increase row by row, initial impact parameters $b_0\/$ increase from left to right; see text for discussion.}
         \label{fig:3biggests}
   \end{figure*}
\changed{To analyze the collision outcomes we investigate the sizes of the biggest fragments in each scenario. Figures~\ref{fig:3biggesth} and~\ref{fig:3biggests} plot the mass-fractions of the three most massive fragments $M_\mathrm{frag}/M_\mathrm{tot}\/$ versus time ($M_\mathrm{tot}=M_\mathrm{P}+M_\mathrm{T}\/$) for the hydro and solid models, respectively. The largest fragment is plotted in red, the next in green and the third-largest in blue. As the fragments are plotted in that sequence it may happen that a smaller fragment hides the dots of a larger one if they are of almost the same size. The data in each frame originate from one initial impact parameter $b_0\/$ from 0 (head-on) to 1,892\,km (fly-by except for the slowest encounter) and initial velocity $v_0\/$. We organized the figures such that each row of sub-plots corresponds to one impact velocity and each column to one initial impact parameter $b_0\/$ as stated on top of the columns. The plots state the actual impact velocity $v_\mathrm{0,i}\/$ as averaged row by row from Tables~\ref{t:actualparsheadon} and~\ref{t:actualparameters}. For non-obvious scenarios the plots also include the impact angles $\alpha\/$ which were taken from Table~\ref{t:actualparameters}.

Most scenarios reflect the picture of Fig.~\ref{fig:20k-250ki950} in that the fragment sizes decrease sharply after the largest one or two fragments. In the following we discuss the plots line by line i.e., by increasing collision velocity. In order to track water we will also refer to Fig.~\ref{fig:waterfraction} which shows how much of the total available water $M_\mathrm{w}=M_\mathrm{T}\times 30\,\%\/$ stays in the three largest surviving fragments after the duration of the simulation.

\paragraph{Impact velocity $v_\mathrm{0,i}\approx 0.95\,v_\mathrm{esc}\/$}~\\ In the hydro model the two bodies merge perfectly for $b_0\le 946\,\mathrm{km}\/$. In the comparable solid scenarios we observe `less perfect' merging indicated by the existence of smaller fragments amounting for negligible mass fractions. A peculiar case is $b_0= 819\,\mathrm{km}\/$ where the bodies merge perfectly in hydro, but the solid model results in a merged body rotating at such a high rate that it loses mass due to the high angular momentum (red curve declining to $\approx 97\,\%\/$ in Fig.~\ref{fig:3biggests}). The material loss also causes the total number of fragments to rise sharply in Fig.~\ref{fig:fragcount}b (top frame, $b_0=819\,\mathrm{km}\/$ curve). The decline in the total water retained in the large fragments which drops from $\approx 100\,\%\/$ ($\alpha=23^\circ\/$) to $\approx 85\,\%\/$ ($\alpha=25^\circ\/$, cf.\ Fig.~\ref{fig:waterfraction}b) indicates a disproportionately high loss of water.

For the largest impact parameter ($b_0=1892\,\mathrm{km}\/$) only this velocity leads to a collision: While the final outcome is very similar for hydro and solid -- one survivor with $>90\,\%\,M_\mathrm{tot}\/$ -- the paths there differ. Due to the relatively low velocity just below the escape velocity and high collision angle $\alpha\approx 60^\circ\/$ the bodies merge after the impact, but rotate quickly. In the hydro model some mass is lost due to to the high angular momentum with the main body staying intact as a spheroid that loses light material (water ice, cf.\ Fig.~\ref{fig:waterfraction}a at $\alpha\approx 60^\circ\/$, $v_\mathrm{0,i}= 0.94\,v_\mathrm{esc}\/$) while it rotates and approaches spherical symmetry. The solid model predicts a more complex merging pattern: a material bridge that breaks up after one ``dumbbell-revolution'' followed by the two primary fragments colliding again to finally form a rotating spheroid that loses icy and rocky material (and re-accretes some of it) while it approaches a spherical shape. The reason for this different behavior is not necessarily found in the physical model though since the actual impact conditions differ slightly (see Table~\ref{t:actualparameters}): While the impact angles match within their respective uncertainties ($58\pm 5\/$\,degrees in hydro versus $62\pm 5\/$\,degrees in solid) the solid model collision velocity $0.99\,v_\mathrm{esc}\/$ is higher than in the hydro model ($0.95\,v_\mathrm{esc}\/$). While Fig.~\ref{fig:fragcount} reflects this behavior by an increasing fragment count for both hydro and solid ($v_\mathrm{0,i}\approx 0.95\,v_\mathrm{esc}\/$, $b_0=1,892\,\mathrm{km}\/$), Fig.~\ref{fig:i1900v231} shows snapshots of this collision.
   \begin{figure*}
            \includegraphics[width=\linewidth, clip]{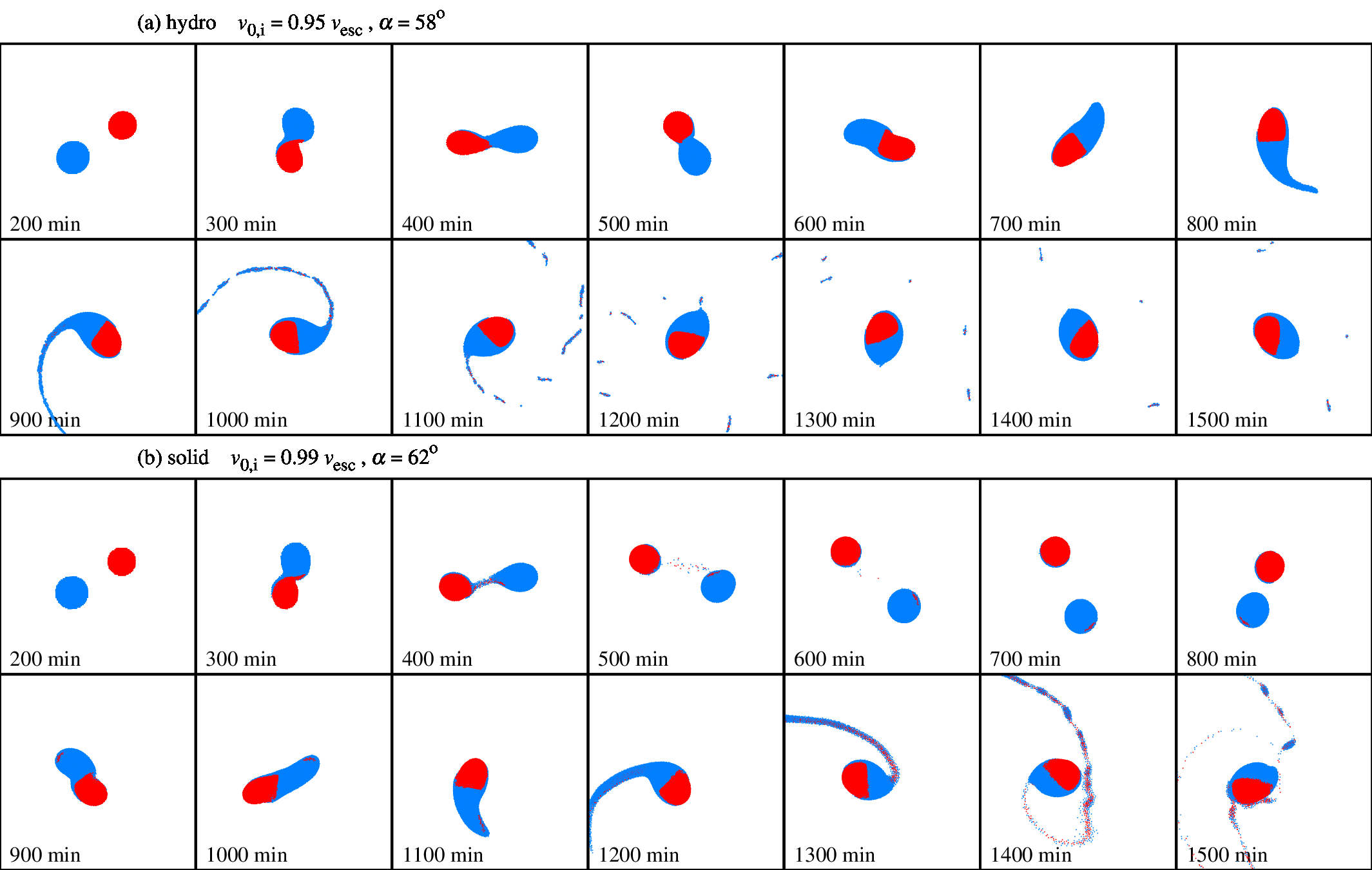}
      \caption{Snapshots of the $b_0=1892\/$\,km, $v_0=231\/$\,m/s scenario, $v_\mathrm{0,i}\/$ and $\alpha\/$ stand for the actual collision velocity and impact angle in (a) the hydro and (b) the solid model. The time stamp indicates elapsed minutes since simulation start, red and blue dots represent basalt and water ice particles, respectively. The size of each frame is $6,000\,\mathrm{km}\times 6,000\,\mathrm{km}\/$ ($xy\/$-projection).}
         \label{fig:i1900v231}
   \end{figure*}

For all higher initial velocities at this impact parameter $b_0=1,892\,\mathrm{km}\/$ the initial separation is large enough for the planetesimals to feel tidal forces but retain enough momentum to escape on hyperbolic orbits around their barycenter (fly-by). We will not discuss these fly-by scenarios further.

\paragraph{Impact velocity $v_\mathrm{0,i}\approx 1.32\,v_\mathrm{esc}\/$}~\\ Again there is perfect merging in the hydro model for small impact parameters ($b_0\le 437\,\mathrm{km}\/$) and ``near perfect'' merging in the solid model.

In the $b_0=819\,\mathrm{km}\/$ case illustrated in Fig.~\ref{fig:i800v500} both models result in a material bridge forming which breaks up -- indicated in Figs.~\ref{fig:3biggesth} and~\ref{fig:3biggests} by the time period in which there are two main fragments of about half the total mass each. At at a time of roughly 500\,minutes into the simulation the two bodies merge as sufficient linear kinetic energy has been converted into rotational and internal deformation energy, which is consistent with prior SPH-based simulations of strengthless larger bodies (planetary embryos) by \citet{agnasp04}. As indicated in the collision snapshots in Fig.~\ref{fig:i800v500} the resulting dumbbell-like body rotates quickly and loses primarily water while approaching a spherical shape and a mass of $\approx 95\,\%\,M_\mathrm{tot}\/$. From about 1000\,min into the simulation we also observe a significantly increasing fragment count originating from material loss (Fig.~\ref{fig:fragcount}, $v_\mathrm{0,i}\approx 1.31\,v_\mathrm{esc}\/$, $b_0=819\,\mathrm{km}\/$). The water loss is seen most easily in Fig.~\ref{fig:waterfraction}: compared to lower impact angles for this encounter velocity, the water on the three biggest fragments decreases from $\approx 100\,\%\/$ to $\approx 82\,\%\/$ of the total water in the system.
   \begin{figure*}
            \includegraphics[width=\linewidth, clip]{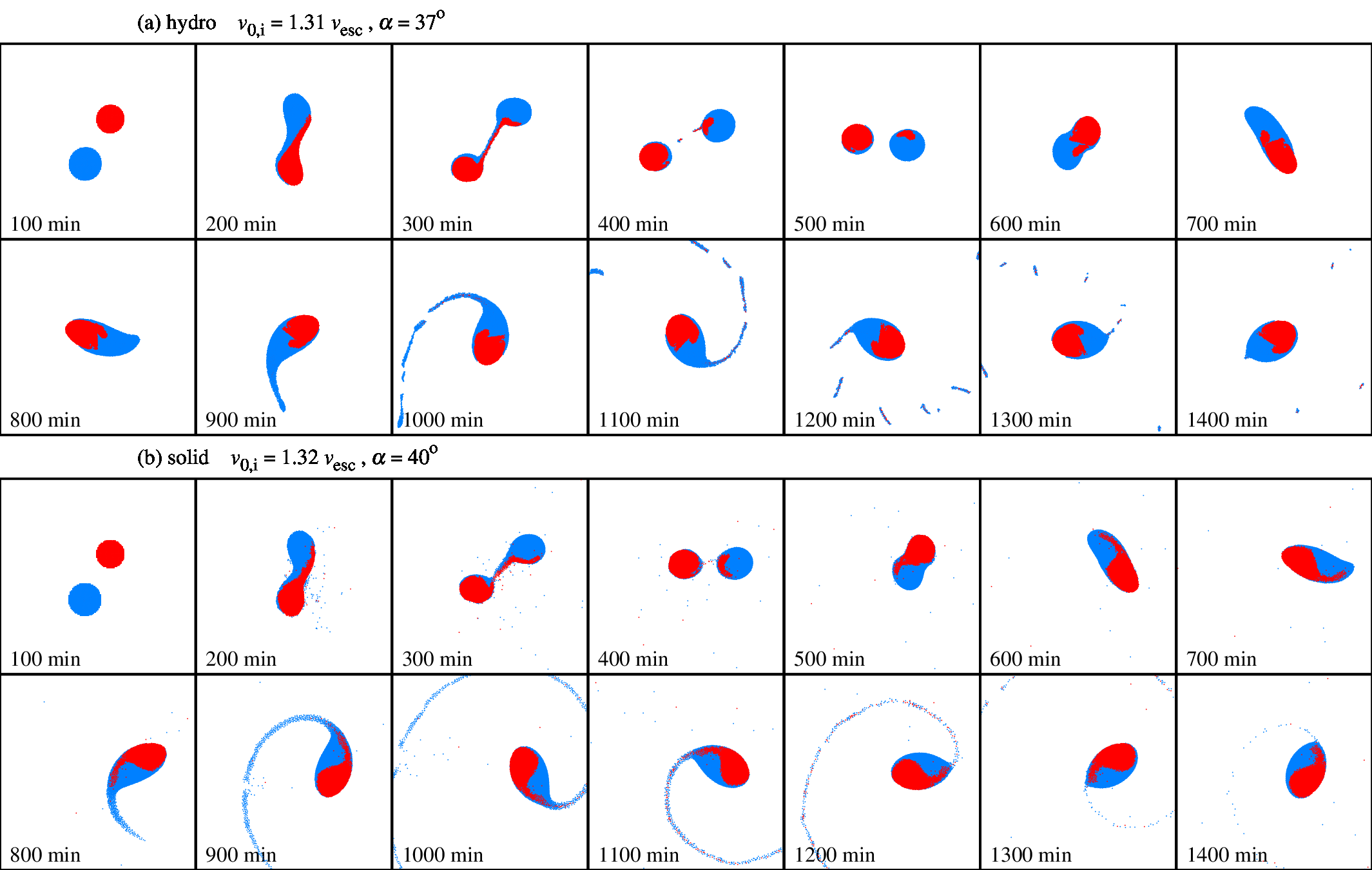}
      \caption{Snapshots of the $b_0=819\/$\,km, $v_0=516\/$\,m/s scenario, $v_\mathrm{0,i}\/$ and $\alpha\/$ stand for the actual collision velocity and impact angle in (a) the hydro and (b) the solid model. The time stamp indicates elapsed minutes since simulation start, red and blue dots represent basalt and water ice particles, respectively. The size of each frame is $6,000\,\mathrm{km}\times 6,000\,\mathrm{km}\/$ ($xy\/$-projection).}
         \label{fig:i800v500}
   \end{figure*}

A further increase in the initial impact parameter $b_0=946\,\mathrm{km}\/$, $\alpha\gtrsim 44^\circ\/$ results in a hit-and-run encounter: the planetesimals stay together shortly and the formation and subsequent break-up of a material bridge leads to two about equally large fragments escaping, strongly deflecting each other's orbits.

\paragraph{Impact velocity $v_\mathrm{0,i}\approx 1.35\,v_\mathrm{esc}\/$}~\\ Similar to the case of smaller impact velocities, up to $b_0\le 437\,\mathrm{km}\/$ the bodies merge in both the hydro and solid scenarios. The solid model indicates some mass loss into smaller fragments after the collision which eventually are re-accreted by the merged body (little dips in the mass fraction of the largest fragment in Fig.~\ref{fig:3biggests}).

In the hydro model the $b_0=819\,\mathrm{km}\/$ case results in a very similar behavior to $v_\mathrm{0,i}\approx 1.32\,v_\mathrm{esc}\/$: mass loss and increasing fragmentation (cf.\ Fig.~\ref{fig:fragcount}a); again, a relatively high amount of water is lost (cf.\ $\alpha=38^\circ\/$, $v_\mathrm{0,i}= 1.36\,v_\mathrm{esc}\/$ in Fig.~\ref{fig:waterfraction}a). However, the resulting configuration is completely different in the solid model (see Fig.~\ref{fig:i800v550}): the bridge that forms right after the collision breaks apart and the two bodies escape after causing two events of fragmentation followed by some re-accretion as seen in Fig.~\ref{fig:fragcount}b at $v_\mathrm{0,i}= 1.36\,v_\mathrm{esc}\/$, $b_0=819\,\mathrm{km}\/$. This marks the onset of hit-and-run at $\alpha=40^\circ\/$ and $v_{0,\mathrm{i}}\/$ between 1.32 and $1.35\,v_\mathrm{esc}\/$ and is in agreement with findings on giant collisions obtained via strengthless-model SPH calculations, see Fig.~17 of \citet{asp09} that shows the transition between merging and hit-and-run at $v_{0,\mathrm{i}}\approx 1.41\,v_\mathrm{esc}\/$ for $\alpha=30^\circ\/$ and $v_{0,\mathrm{i}}\approx 1.12\,v_\mathrm{esc}\/$ for $\alpha=45^\circ\/$. Additionally, in Fig.~\ref{fig:i800v550} we observe some water being transferred to the target (which we do not investigate further at this point).
   \begin{figure*}
            \includegraphics[width=\linewidth, clip]{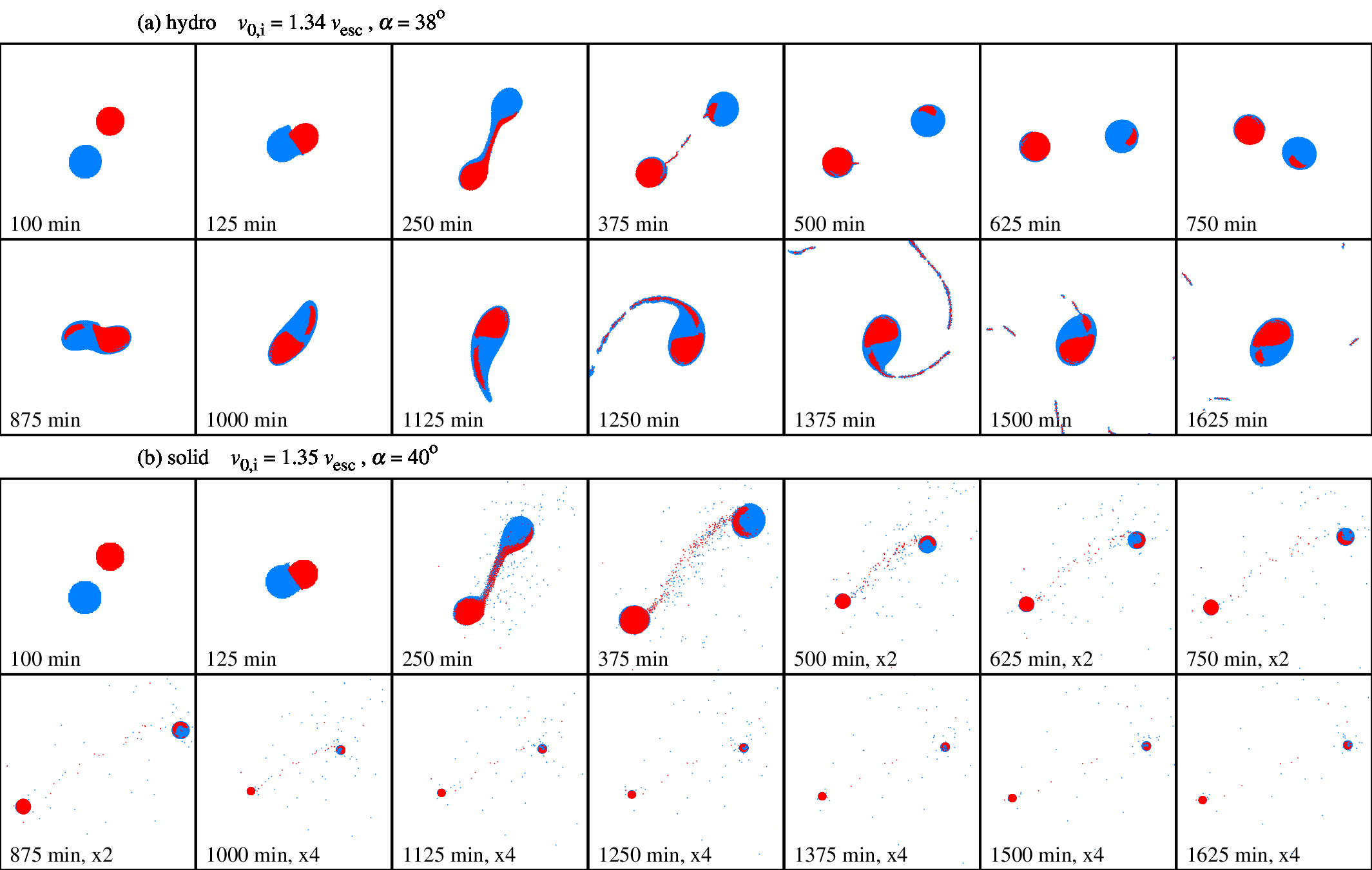}
      \caption{Snapshots of the $b_0=819\/$\,km, $v_0=537\/$\,m/s scenario, $v_\mathrm{0,i}\/$ and $\alpha\/$ stand for the actual collision velocity and impact angle in (a) the hydro and (b) the solid model. The time stamp indicates elapsed minutes since simulation start along with a zoom-out factor if applicable; red and blue dots represent basalt and water ice particles, respectively. The size of each frame varies from $6,000\,\mathrm{km}\times 6,000\,\mathrm{km}\/$ to $24,000\,\mathrm{km}\times 24,000\,\mathrm{km}\/$ depending on the zoom-out factor ($xy\/$-projection).}
         \label{fig:i800v550}
   \end{figure*}

Increasing the initial impact parameter to $b_0=946\,\mathrm{km}\/$ shows practically the same behavior as in the slightly slower encounter ($v_\mathrm{0,i}\approx 1.32\,v_\mathrm{esc}\/$): the objects merge for a short period, a material bridge forms, breaks, and finally two fragments of about $0.5\,M_\mathrm{tot}\/$ each escape.

\paragraph{Impact velocity $v_\mathrm{0,i}=2.12\,v_\mathrm{esc}\/$}~\\
For $b_0\le 219\,\mathrm{km}\/$ ($\alpha \le 12^\circ\/$) we still observe merging, but in both models it involves some erosion shortly after the impact and most of the eroded material being re-accreted by the one surviving body. Erosion is more prominent in the solid model ($\gtrsim 10\,\%\,M_\mathrm{tot}\/$ initially, $>90\,\%\,M_\mathrm{tot}\/$ in the survivor) than in the hydro case ($\lesssim 5\,\%\,M_\mathrm{tot}\/$ initial fragmentation and almost $100\,\%\,M_\mathrm{tot}\/$ in the survivor). Also, the solid model predicts higher loss of water -- less than 70\,\% of the water remains in the surviving fragment as opposed to $>95\,\%\/$ in hydro (see Fig.~\ref{fig:waterfraction}).

Higher impact angles lead to a material bridge between the planetesimals which eventually breaks up and two to three bodies are formed. Again, the solid model predicts more water loss than the hydro model. In the latter the final outcome for $24^\circ\le \alpha\le 59^\circ\/$ is a hit-and-run scenario (two surviving bodies of about equal mass). In the solid case we observe three main fragments -- two between 45 and $50\,\%\,M_\mathrm{tot}\/$ and one about $3\,\%\,M_\mathrm{tot}\/$ -- for $\alpha=25^\circ\/$, two about equally massive ones for $\alpha=50^\circ\/$, and three surviving bodies of about 50, 40, and 5\,\% of $M_\mathrm{tot}\/$ for $\alpha=62^\circ\/$, respectively.

\paragraph{Impact velocity $v_\mathrm{0,i}\approx 3.04\,v_\mathrm{esc}\/$}~\\ For central and near-central impacts ($b_0\le 219\,\mathrm{km}\/$, $\alpha \le 12^\circ\/$) the model results differ: while in the hydro case most of the mass remains in few large fragments the solid model predicts a large amount of the material being lost by a very high degree of fragmentation: a central impact results in one survivor of noteworthy mass in both models, but it amounts for about 95\,\% of $M_\mathrm{tot}\/$ in hydro and for only about 60\,\%\,$M_\mathrm{tot}\/$ in solid. Both models predict water loss, but again more in the solid case (cf.\ Fig~\ref{fig:waterfraction}; this also holds for larger collision angles). Increasing the collision angle to $\alpha = 12^\circ\/$ gives two survivors of just under half the total mass in hydro (after a quite complex pattern of fragmentation and re-accretion) and (at least) three survivors in the $15-25\,\%\,M_\mathrm{tot}\/$ range in the solid model. Figure~\ref{fig:i200v1500}a shows multiple fragments being re-accreted by the two main survivors in the hydro case whereas there are multiple surviving fragments for the whole duration of the simulation in the solid model (beyond the time span shown in Fig.~\ref{fig:i200v1500}).

The outcome of the remaining collision angles look very similar in both models: there are two main survivors of around 40\,\%\,$M_\mathrm{tot}\/$ for $\alpha = 25^\circ\/$ and 50\,\%\,$M_\mathrm{tot}\/$ for larger impact angles.
   \begin{figure*}
            \includegraphics[width=\linewidth, clip]{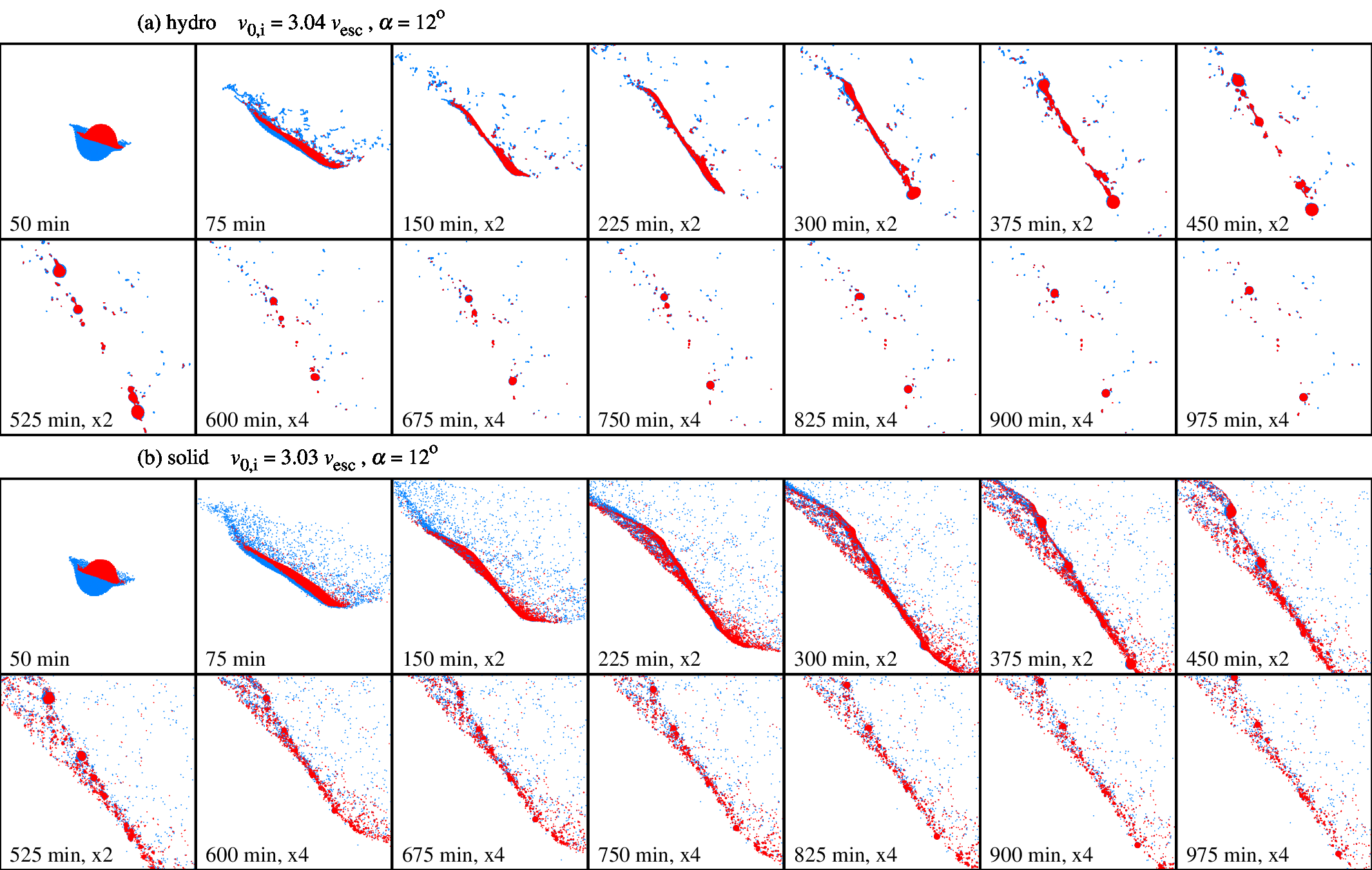}
      \caption{Snapshots of the $b_0=219\/$\,km, $v_0=1.5\/$\,km/s scenario, $v_\mathrm{0,i}\/$ and $\alpha\/$ stand for the actual collision velocity and impact angle in (a) the hydro and (b) the solid model. The time stamp indicates elapsed minutes since simulation start along with a zoom-out factor if applicable; red and blue dots represent basalt and water ice particles, respectively. The size of each frame varies from $6,000\,\mathrm{km}\times 6,000\,\mathrm{km}\/$ to $24,000\,\mathrm{km}\times 24,000\,\mathrm{km}\/$ depending on the zoom-out factor ($xy\/$-projection).}
         \label{fig:i200v1500}
   \end{figure*}

\paragraph{Impact velocity $v_\mathrm{0,i}\ge 3.97\,v_\mathrm{esc}\/$}~\\ The collision outcomes are very similar between the hydro and solid models with a tendency towards smaller survivors in the solid case due to the significantly higher degree of fragmentation. Central impacts lead to one $\approx 30\,\%\,M_\mathrm{tot}\/$ survivor ($v_\mathrm{0,i}=3.97\,v_\mathrm{esc}\/$) and an about 10\,\% (hydro) / 15\,\% (solid) $M_\mathrm{tot}\/$ survivor after the faster impact ($v_\mathrm{0,i}=5.88\,v_\mathrm{esc}\/$). The hydro model predicts two small bodies of $\lesssim 5\,\%\,M_\mathrm{tot}\/$ in the latter case that are not there in solid ($\ll 5\,\%\,M_\mathrm{tot}\/$).

Increasing the impact angle ($\alpha = 13^\circ\/$) leads to increasing erosion: three survivors in the $v_\mathrm{0,i}=3.97\,v_\mathrm{esc}\/$ case ($\lesssim 15\,\%\,M_\mathrm{tot}\/$ hydro, $\lesssim 10\,\%\,M_\mathrm{tot}\/$ solid) and extensive disruption for $v_\mathrm{0,i}=5.88\,v_\mathrm{esc}\/$ in both models. The next collision angle increment ($\alpha = 27/28^\circ\/$) leads to two survivors of just under 40\,\% (hydro) and 30\,\% (solid) of $M_\mathrm{tot}\/$ for $v_\mathrm{0,i}=3.97\,v_\mathrm{esc}\/$. Increasing the collision speed to $5.88\,v_\mathrm{esc}\/$ virtually disrupts the bodies in the solid model, but leaves three $\lesssim 15\,\%\,M_\mathrm{tot}\/$ fragments in hydro. A further increase of the collision angle results in two surviving bodies of about 50\,\%\,$M_\mathrm{tot}\/$ (hit-and-run). Comparing the water retention in Fig.~\ref{fig:waterfraction} and the fragment sizes reveals similar water loss in the hydro and solid models.

Globally the figures discussed above tell that there is -- dependent on the 
velocity and the encounter angle -- a difference in the outcome for the hydro and solid models for some of the 
collisions. Especially low-energy collisions and near-central impacts at medium energy deserve a closer investigation regarding the best fitted physical model. It also becomes evident that -- we stressed it before -- most scenarios deserve a separate computation with parameters such as resolution, initial SPH particle distribution, integration period, algorithm parameters, etc.\ tailored to fit the key questions to be answered.
}

   \begin{figure}
            \includegraphics[width=\linewidth, clip]{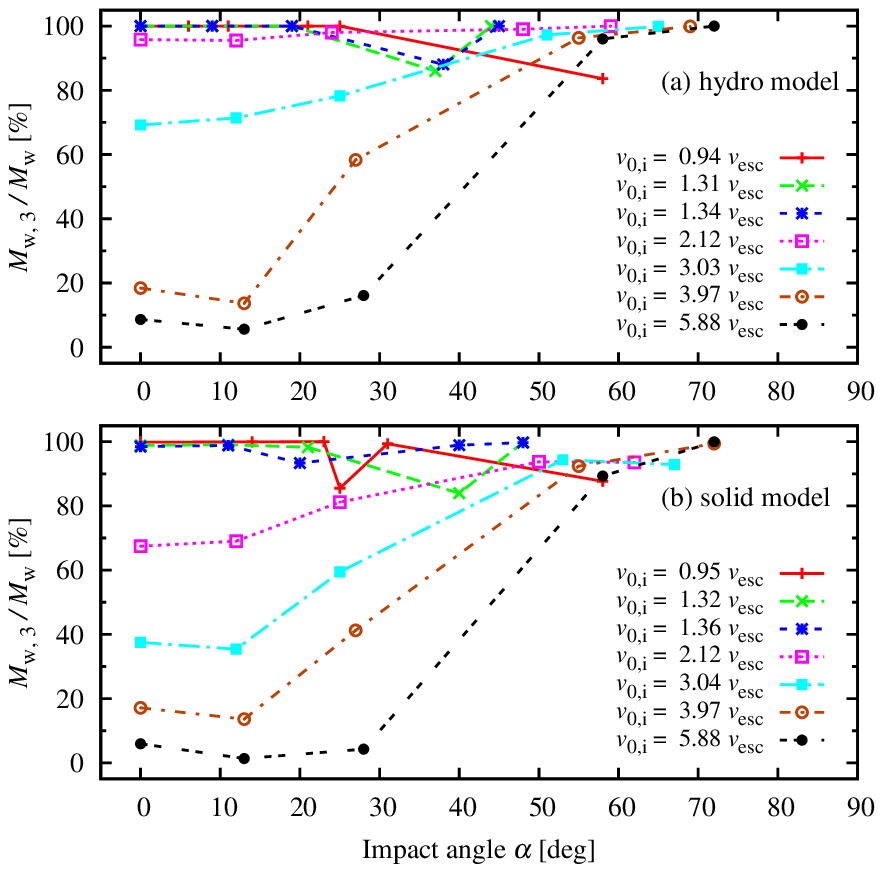}
      \caption{Water remaining in the three biggest fragments $M_\mathrm{w,3}\/$ as a percentage of the total water in the system $M_\mathrm{w}\/$ after the collision (2000\,min after simulation start).
              }
         \label{fig:waterfraction}
   \end{figure}

\section{Conclusions and future research}
\label{sect:conclusion}

We established qualitative differences between collision outcomes obtained by the solid and the hydro models. In general, the solid models predict significantly higher numbers of fragments and dust cloud-like ejecta dispersed over a much greater volume than the collision debris in the hydro case. \changed{While the collision outcome is similar for many investigated scenarios in terms of number and sizes of the surviving bodies, the loss of material -- especially water (ice) -- is consistently bigger in the solid model.}

\changed{Additionally, our simulations indicate that collisions characterized by parameters as they are found by dynamical n-body simulations of early planetary systems can transfer water (ice) between Ceres-sized bodies. An examination of the fate of water on the hit-and-run survivors is left for future studies.}

As the outcome of the collisions show similarities with existing giant impact results obtained via strengthless solid body models (cf.\ Sect.~\ref{sect:scenres}) we feel to have demonstrated that there is need for more detailed quantitative studies to (a) verify our observations regarding the amount of material in general and water (ice) in particular transferred and lost during planetesimal collisions and (b) put them in a quantitative context. For this fragments need to be investigated with respect to their masses and energies determining whether they are bound to the system of colliding bodies or whether they escape either indefinitely or beyond the system's Hill radius. As we observe more fragments -- also larger ones -- in the solid model analyzing the fragment distribution will most likely contribute to investigating the origin of asteroid families from dynamical \citep[cf.\ ][]{galsou11} and size statistics \citep[cf.\ ][]{knemil03} perspectives.

\changed{An important question these quantitative investigations will answer is whether the difference between the hydro and solid models is large enough to justify the use of solid models when simulating planetesimal collisions in dynamic studies. We identified low energy configurations and mid-energy collisions with small impact parameter as candidate scenarios for solid model simulations. These are significantly more expensive from a computational point of view which will make the choice of the right model a practical issue if considering a ``collision outcome catalog''. Such a catalog can subsequently be incorporated in n-body dynamical studies of early planetary systems and will augment using fitted formulas for giant collision outcomes \citep[cf.\ ][]{genkok12}. The large parameter space that such a catalog will have to cover still requires thousands of collision simulations which we plan to tackle deploying a high-performance GPU code. First numerical experiments suggest a speedup by a factor of about 50 compared to a parallel CPU implementation \citep{rie14}.}

As we have seen, off-center impacts result in rotating survivors. While there is some indication that more initial linear kinetic energy is converted into internal and/or rotational energy in the hydro model than it is the case in the solid model, more detailed quantitative studies are necessary. Part of these investigations will be studying collision outcomes of initially rotating bodies.

\begin{acknowledgements}
This research is produced as part of the FWF Austrian Science Fund project S~11603-N16. \changed{This publication was supported by FWF. In part the calculations for this work were performed on the hpc-bw-cluster -- we gratefully thank the bwGRiD project\footnote{\tiny bwGRiD (http://www.bw-grid.de), member of the German D-Grid initiative, funded by the Ministry for Education and Research (Bundesministerium fuer Bildung und Forschung) and the Ministry for Science, Research and Arts Baden-Wuerttemberg (Ministerium fuer Wissenschaft, Forschung und Kunst Baden-Wuerttemberg).} for the computational resources. We also thank the anonymous referee for many comments and suggestions which significantly improved this study.}
\end{acknowledgements}


\bibliographystyle{aa} 
\bibliography{references} 

\end{document}